\documentclass{aa}

\usepackage{graphicx}
\usepackage{txfonts}
\usepackage{natbib}
\usepackage{lipsum}

\usepackage[usenames,dvipsnames]{xcolor}
\usepackage{url}
\bibpunct{(}{)}{;}{a}{}{,}

\newcommand{\be}{\begin{equation}}
\newcommand{\ee}{\end{equation}}

\def\apjl{ApJL}
\def\apjs{ApJS}

\newcommand{\Mpc}{$h^{-1}$\thinspace Mpc}

\newcommand{\vmh}{h^{-1}\mathrm{Mpc} }

\DeclareMathAlphabet{\pazocal}{OMS}{zplm}{m}{n}

\usepackage{amstext}

\begin{document}  

\title{
The richest clusters in the Coma and Leo superclusters:\\
Properties and evolution
} 

\author {Maret~Einasto\inst{1} 
\and Peeter~Tenjes\inst{1} 
\and Rain~Kipper\inst{1} 
\and Pekka~Hein\"am\"aki\inst{2}
\and Elmo~Tempel\inst{1,3} 
\and Lauri Juhan~Liivam\"agi\inst{1} 
\and Michael J. West\inst{4}
\and Boris Deshev\inst{1} 
\and Jaan~Einasto\inst{1,3,5}
}
\institute{Tartu Observatory, University of Tartu, Observatooriumi 1, 61602 T\~oravere, Estonia
\and 
Tuorla Observatory, Department of Physics and Astronomy, University of Turku, 20014
Turku, Finland
\and
Estonian Academy of Sciences, Kohtu 6, 10130 Tallinn, Estonia
\and
Lowell Observatory, 1400 W. Mars Hill Rd., Flagstaff, AZ, 86001 USA
\and
ICRANet, Piazza della Repubblica 10, 65122 Pescara, Italy
}

\authorrunning{Einasto, M. et al. }

\date{ Received   / Accepted   }

\titlerunning{Coma-Leo}

\abstract
{
Superclusters of galaxies represent
dynamically active environments in which galaxies and their systems form and evolve. 
}
{We study the substructure, connectivity, and galaxy content 
of galaxy clusters \object{A1656} and \object{A1367} in the Coma supercluster and of \object{A1185}
in the Leo supercluster 
with the aim of understanding the evolution of clusters from turnaround to virialisation, and 
the evolution of whole superclusters.
}
{
We used data from the Sloan Digital Sky Survey DR10 MAIN galaxy sample and from DESI 
cluster catalogues.
The projected phase space diagram and the distribution of mass 
were used to identify regions of various infall stages (early and late infall, 
and regions of ongoing infall, i.e. regions of influence), 
their characteristic radii, embedded mass, and
density contrasts in order to study the evolution of
clusters 
with 
the spherical collapse model. We determined the substructure of clusters 
using normal mixture modelling and their connectivity by counting filaments 
in the cluster's regions of influence.
We analysed galaxy content in clusters and in their environment and 
derived scaling relations between cluster masses.
}
{All three clusters have a substructure with two to five components and 
up to six filaments connected to them. The radii of regions of influence are 
$R_\mathrm{inf} \approx 4$~\Mpc, and the density contrast at their borders is  $\Delta\rho_{inf} \approx 50 - 60$.
The scaling relations between the masses of clusters have a very small scatter. 
The galaxy content of the clusters and of their regions of influence vary from
cluster to cluster. In high-density regions (superclusters), the percentage of 
quiescent galaxies is higher than in low-density regions between superclusters,
where approximately one-fourth of the galaxies are still quiescent. 
}
{
The collapse of the regions of influence of clusters started at redshifts
$z \approx 0.4 - 0.5$. Clusters will be virialised approximately in $\approx 3.3$~Gyrs. 
Clusters in superclusters will not merge, and their present-day turnaround regions will be virialised in
$\approx 10$~Gyrs.
The large variety of properties of clusters suggests that they 
have followed different paths during evolution. 
}

\keywords{large-scale structure of the Universe - 
galaxies: groups: general - galaxies: clusters: general: - galaxies: clusters: individual}

\maketitle

\section{Introduction} 
\label{sect:intro} 

The large-scale distribution of matter in the Universe forms a pattern called the cosmic web -- 
a huge network of galaxies and galaxy groups and clusters connected by galaxy filaments and 
separated by huge underdense regions where the density of matter is clearly lower
\citep{1978MNRAS.185..357J, 1988Natur.334..129K}.
Among the various structures of the cosmic web, rich galaxy clusters deserve special attention. 
Rich clusters are the largest
objects in the cosmic web that can currently be virialised \citep{2012ARA&A..50..353K}.
Galaxy clusters as nodes in the cosmic web grow by infall of galaxies and groups along filaments. 
Simulations have shown that present-day rich clusters with a mass of at least $M_{z0} \approx 10^{14}M_\odot$
have collected their galaxies along filaments from regions with co-moving radii 
of at least $10$~\Mpc\ \citep{2013ApJ...779..127C}.  
These regions around clusters are referred to as the spheres of dynamical influence 
of the clusters, and in these areas, all galaxies, groups, and filaments are 
falling into clusters \citep{2020A&A...641A.172E, 2021A&A...649A..51E}.
The sizes of the regions of the dynamical influence depend on the cluster mass and 
on the assembly history of the infalling systems \citep{2013ApJ...779..127C, 2013MNRAS.430.3017B}. 
Simulations have demonstrated that the properties of clusters and 
the properties of their close environment are related; more massive clusters 
also have a higher number of substructures and higher connectivity (number of filaments connected 
to a cluster)
\citep{2018MNRAS.479..973C, 2021A&A...651A..56G, 2025A&A...700A.182B}. 

These results have been confirmed by observational studies. 
All rich clusters in the nearby Universe lie in 
filaments in superclusters or in supercluster outskirts. 
More massive and richer clusters tend to have a more complicated or clumpy structure,
characterised by higher multimodality (a larger number of substructures), and they contain a higher fraction of 
quiescent galaxies than less massive clusters. 
In the spheres of influence of clusters, the connectivity is higher for rich clusters
\citep[][and references therein]{2012A&A...540A.123E, 2021A&A...651A..56G, 2022A&A...668A..69E,
2024A&A...681A..91E}. 

Simultaneous studies of substructure, connectivity, and the galaxy content of clusters and their spheres of influence from
observations have only been
done for a small number of clusters, namely, the richest
clusters in A2142 and in the Corona Borealis superclusters
\citep{2020A&A...641A.172E, 2021A&A...649A..51E}. 
One aim of this study is to extend the sample of clusters for 
which substructure, connectivity, and the galaxy content are analysed together.
We analysed the properties and galaxy content of
the richest clusters in the Coma and Leo superclusters, namely,
the  low $z\simeq 0.03$ redshift Coma (Abell cluster \object{A1656}),
\object{A1367}, and \object{A1185}  clusters (see Table~\ref{tab:cl}). 
These clusters have been studied extensively, and this makes them ideal testbeds for comparing various 
methods and results \citep[see, for example, ][and references therein]{2020MNRAS.497..466S, 2020A&A...634A..30M, 2025A&A...694A.216J}.

The evolution of a spherical  perturbation in an expanding universe, 
which becomes a rich cluster in the present-day Universe, is usually described by 
applying the spherical collapse model \citep{1980lssu.book.....P,
1984ApJ...284..439P, 1991MNRAS.251..128L}.  
In this model, the evolution of a spherical shell is determined by the 
mass in its interior. At first, overdensities in the cosmic web expand as the universe expands.
If the mass within the overdensity is sufficiently high (as in progenitors of 
present-day rich clusters), at a certain moment the expansion stops and the collapse begins. This epoch is called the
turnaround. The final stage of the collapse is called virialisation. Overdensities in which the density contrast at 
present is not high enough for turnaround may reach turnaround in the future. Therefore, the evolution of
a spherical overdensity is characterised by the following important epochs
and corresponding characteristic density contrasts: turnaround, future collapse, and virialisation
\citep{2015A&A...575L..14C, 2015A&A...581A.135G}. 
In our study, we calculate the mass (density) distribution around clusters to find masses and radii at
characteristic density contrasts and find redshifts of the 
main epochs for these three clusters. These redshifts are used to discuss their  possible evolution.

These epochs are related to the various stages of the infall of galaxies to clusters
(early and late infall of galaxies and regions of ongoing infall where galaxies
are  falling into clusters, i.e. regions of influence of clusters),
which can be studied with the projected phase space (PPS) diagram.
Using the PPS diagram, \citet{2020A&A...641A.172E} and  \citet{2021A&A...649A..51E}
detected several density minima in the distribution of galaxies and galaxy groups 
around the richest galaxy clusters, which mark different infall regions in and around clusters with 
their characteristic radii;
cluster radius, $R_\mathrm{cl}$, and the radius of the 
sphere of influence around clusters, $R_\mathrm{inf}$. 
We applied the PPS diagram to find these regions and to determine the corresponding radii for clusters
under study. To determine the connectivity of clusters, 
we searched for filaments within the spheres of influence.

We determined the substructure of clusters in two different ways. 
First, we applied a normal mixture model to determine the substructure of clusters, 
as this was done, for example, in \citet{2012A&A...540A.123E}. 
Second, we compared the substructure of clusters determined in this way with DESI clusters presented 
in a merging cluster catalogue by \citet{2024MNRAS.532.1849W}.
Then we
compared the masses, substructure, connectivity, various characteristic 
radii, and galaxy content of clusters.
Usually, it is considered that galaxies in 
groups and clusters are mostly  passive, 
with old stellar populations,
while galaxies falling into clusters for the first time are either star forming or were 
preprocessed in groups before infall into clusters 
\citep[see, for example][for a brief review and references]{2013MNRAS.430.3017B}. 
Therefore, the study of the galaxy content of components in clusters together with their location in the PPS 
diagram as well as the study of galaxy content in a larger-scale environment around
clusters may shed light on the  infall history of structures in and near clusters.

Our main dataset is the Sloan Digital Sky Survey (SDSS) DR10 MAIN spectroscopic 
galaxy sample in the redshift range $0.009 \leq z \leq 0.200$.
SDSS data are used to calculate the luminosity-density
field of galaxies, to determine groups and filaments in the galaxy distribution,
and to get data on galaxy properties
\citep{2011ApJS..193...29A, 2014ApJS..211...17A}.
The luminosity-density field with a smoothing length of $8$~\Mpc, $D8$,
characterises the global environment of galaxies and was used to determine supercluster member galaxies and groups.

In addition to the SDSS data, we also used data from
the catalogue of galaxy clusters based on photometrical data of galaxies from the 
Dark Energy Spectroscopic Instrument (DESI) Legacy Survey DR10 
\citep{2024AJ....168...58D}\footnote{https://www.legacysurvey.org/}
by \citet{2024ApJS..272...39W} (hereafter WH24). 
On the basis of this catalogue, \citet{2024MNRAS.532.1849W}
compiled a catalogue of merging galaxy clusters and subclusters.  We additionally
used the data from this catalogue.
We  identified rich clusters in the Coma-Leo region in all of these  catalogues and  compared  the substructure 
of clusters based on various data.
In accordance with the studies by \citet{2020A&A...641A.172E} and \citet{2021A&A...649A..51E}  
on rich galaxy clusters in superclusters,
we applied the following cosmological parameters: the Hubble parameter $H_0=100~ 
h$ km~s$^{-1}$ Mpc$^{-1}$, the matter density $\Omega_{\rm m} = 0.27$, and 
the dark energy density $\Omega_{\Lambda} = 0.73$ 
\citep{2011ApJS..192...18K}.

\section{Data} 
\label{sect:data} 

{\emph{Superclusters}} in the Coma--Leo region were identified in the supercluster catalogue 
by \citet{2012A&A...539A..80L}, 
based on
the luminosity-density field with smoothing length $8$~\Mpc, $D8$ 
\citep[in units of mean luminosity--density, $\ell_{\mathrm{mean}}$ = 
1.65$\cdot10^{-2}$ $\frac{10^{10} h^{-2} L_\odot}{(\vmh)^3}$ ][]{2012A&A...539A..80L}.
In the luminosity-density field,
connected regions with the
luminosity-density above a threshold density of $D8 = 5.0$
can be defined as  superclusters. 
Underdense regions between superclusters can be divided as supercluster outskirts with 
$1 \leq D8 \leq 5.0$, and extreme underdense regions (voids or watershed regions) with 
$D8 \leq 1.0$  \citep[see also][for this division]{2022A&A...668A..69E, 2024A&A...681A..91E}. 
Luminosity-density field 
characterises the global environment of galaxies and can be used to determine supercluster member 
galaxies and groups.
Our final dataset, which includes  the Coma and the Leo 
superclusters, is  from the redshift range $0.02 \leq z \leq 0.05$,
and sky coordinate range $160^{\degr} \leq R.A. \leq 210^{\degr}$, and $15^{\degr} \leq Dec. \leq 34^{\degr}$.

{\emph {Groups and clusters of galaxies: SDSS sample.}}
We used data from cluster and group catalogues by 
\citet{2014A&A...566A...1T}, based on the SDSS DR10 MAIN spectroscopic 
galaxy sample with 
Galactic extinction-corrected apparent $r$ magnitudes $r \leq 
17.77$ and redshifts $0.009 \leq z \leq 0.200$
\citep{2011ApJS..193...29A, 2014ApJS..211...17A}.
Corresponding group catalogues are available at the CDS \footnote{cdsarc.u-strasbg.fr}.
Groups have been determined by applying the friends-of-friends (FoF) clustering analysis 
method \citep{1982Natur.300..407Z, 1982ApJ...257..423H}. 
Galaxies without any close neighbours are classified as single galaxies which 
may be the brightest galaxies of faint groups where
other group members are too faint to be included in the  SDSS spectroscopic sample,
or these galaxies may also be systems in which one luminous galaxy is surrounded
by dwarf satellites. We  focused on the richest clusters in the Coma and Leo superclusters,
A1656, A1367, and A1185. For comparison, we used data also on the richest clusters in the Corona Borealis and A2142 superclusters,
A2065, A2061, A2089, Gr2064, and A2142 \citep{2020A&A...641A.172E, 2021A&A...649A..51E}. 

{\emph{Groups and clusters of galaxies: DESI sample.}}
WH24 catalogue of galaxy clusters is based on galaxy data from the DESI Legacy Survey DR10. 
From this catalogue we extracted data on galaxy clusters
which lie in the Coma-Leo region, in total 29 clusters with at least six member galaxies. 
WH24 cluster catalogue is based on photometric redshifts of galaxies, but, if available, 
the redshifts of the brightest cluster galaxies were used as the cluster redshifts. 
All clusters in the Coma-Leo region from WH24 have spectroscopic redshifts. 

We present data on the richest clusters 
in the Coma-Leo superclusters region for both samples in Tables~\ref{tab:cl} and \ref{tab:wh24cl}, 
respectively.
These tables provide data on cluster coordinates, richness, luminosity and mass.
In Table~\ref{tab:cl} mass of the cluster is taken from \citet{2014A&A...566A...1T}, and it is calculated assuming the Navarro-Frenk-White (NFW) density profile.
In this Table, and in the further text, we used the notation cluster gravitational radius $R_\mathrm{g}$
for the radius sometimes called the dynamical virial radius, in order to avoid confusion
with other  (cosmological) definitions. 
WH24 provide cluster radius $r_{500}$ and mass $M_{500}$, which are the radius and mass 
within which the mean density is 500 times the critical density of the universe. 
It also provides the number of member galaxy candidates within $r_{500}$, $N_\mathrm{gal}$.

We also used the catalogue of merging clusters and subclusters based on WH24
catalogue \citep[hereafter WH24m; ][]{2024MNRAS.532.1849W}. 
In WH24m, partner cluster systems have been defined as clusters which were close enough to each other
to probably be gravitationally bound. The most massive cluster in these systems 
was considered the main cluster, and other cluster(s) were considered 
partner clusters. 
The criteria to be partner systems were small separation in the sky plane,
with maximum difference as $5r_{500}$, and 
a small velocity difference, with the maximum $1500$~km/s (see WH24m for details).  
WH24 and WH24m cluster catalogues are available on the cluster catalogue web page
\footnote{\url{http://zmtt.bao.ac.cn/galaxy\_clusters/}.}

{\emph{Filament membership of galaxies, groups, and clusters}}. 
To detect filaments and their members, we employed data 
from the filament catalogues by \citet{2014MNRAS.438.3465T} and \citet{2016A&C....16...17T},
available at 
\footnote{cdsarc.u-strasbg.fr}.
In these studies, 3D
galaxy filaments were detected by 
applying a marked point process to the SDSS galaxy distribution (Bisous model, and
hereafter Bisous filaments). For each galaxy, a distance from the nearest filament
axis was calculated. A galaxy was considered to be a filament member  
if its distance from the nearest filament axis $D_{fil} \leq 0.5$~\Mpc.

{\emph{Data on star formation properties of galaxies}}.
Data of the galaxy properties were taken from the SDSS DR10 web page  
\footnote{\url{http://skyserver.sdss3.org/dr10/en/help/browser/browser.aspx}}.
In addition, we analysed the star formation properties of galaxies,
using $D_n(4000)$ index and star formation rate $\log \mathrm{SFR}$,
taken from
the MPA-JHU spectroscopic catalogue \citep{2004ApJ...613..898T, 2004MNRAS.351.1151B}.

$\text{The }D_n(4000)$ index 
\citep{1999ApJ...527...54B} is correlated 
with the time passed since the most recent star formation event in a galaxy \citep{2003MNRAS.341...33K}.
According to this 
index galaxies can be divided into classes 
with old stellar populations and galaxies
which are actively forming stars at present. The limiting value $D_n(4000) = 1.55$ corresponds
to the mean age of stellar populations in a galaxy $1.5$~Gyr 
\citep{2003MNRAS.341...33K, 2017A&A...605A...4H}. 
\citet{2003MNRAS.341...33K} also showed that $D_n(4000) \geq 1.75$ corresponds to a mean age 
of stellar populations of about $4$~Gyr (for Solar metallicity) or older 
(for lower metallicities). Galaxies with $D_n(4000) \geq 2.0$ may have stellar populations 
formed even more than $10$~Gyr ago.
Following \citet{2022A&A...668A..69E}, we called the population of 
galaxies with  $D_n(4000) \geq 1.75$ as galaxies with very old (VO) stellar populations.
 Galaxies with $D_{n}(4000) < 1.35$ have  young stellar populations with a mean age of
$\leq 0.6$~Gyr 
\citep{2003MNRAS.341...33K}. 
Using the star formation rate, star-forming and passive (quenched)
galaxies can be divided at $\log \mathrm{SFR} = -0.5$.  Quenched galaxies have $\log \mathrm{SFR} < -0.5$, and for star-forming galaxies $\log \mathrm{SFR} \geq -0.5$. 
Some galaxies with $D_n(4000) \geq 1.75$ have $\log \mathrm{SFR} \geq -0.5$, thus they
may still be forming stars. However, 
such galaxies comprise less than $2$~\% of the galaxies in the sample. Therefore,
the use of limit $D_n(4000) = 1.75$ to separate VO galaxies 
is justified, as was also found in \citet{2022A&A...668A..69E}. 
Galaxies with $D_n(4000) \geq 2.0$ form $\approx 1$~\% of all galaxies in our sample.

\begin{figure}[ht]
\centering
\resizebox{0.44\textwidth}{!}{\includegraphics[angle=0]{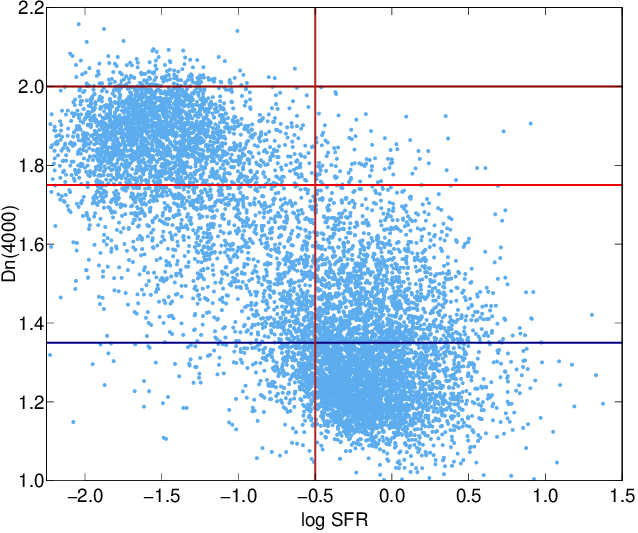}}
\caption{
Star formation rate ($\log \mathrm{SFR}$) versus $D_n(4000)$ index
for galaxies with $ M_r = -18.30$.
Horizontal lines show $D_n(4000)$ limits for galaxies
with stellar populations of different ages, $1.35$, $1.75$, and $2.0$ (see text for details), and the vertical line
shows the limit $\log \mathrm{SFR} = -0.5$ that separates star-forming and quenched galaxies.
}
\label{fig:dn4sf}
\end{figure}

Figure~\ref{fig:dn4sf}, where we plot $\log \mathrm{SFR}$ versus $D_n(4000)$,
shows these limits with horizontal and vertical lines.
These limits are used to study galaxies with various stellar populations in clusters and in surrounding regions. 
At the farthest 
end of our sample, an absolute magnitude limit for a complete sample is $ M_r = -18.30$. 
We used this limit when we compared star formation properties of galaxies in the whole sample 
separately for galaxies in superclusters and in low-density regions around them.

\begin{table*}[ht]
\caption{Data on rich galaxy clusters in the Coma and Leo superclusters based on SDSS DR10.}
\begin{tabular}{rrrrrrrrrrrrrr} 
\hline\hline  
(1)&(2)&(3)&(4)&(5)& (6)&(7)&(8)&(9)&(10)&(11)&(12)&(13)&(14)\\      
\hline 
No. & Abell ID& ID &$N_{\mathrm{gal}}$& $\mathrm{R.A.}$ & $Dec.$ 
&$\mathrm{z}$ &$\sigma_{v}$ &  $R_{\mathrm{g}}$&  $R_{\mathrm{max}}$ & $M_{\mathrm{NFW}}$  
& $L_{\mathrm{tot}}$ &$M/L$ & $D8$ \\
\hline       
 1 &A1656&   52 & 680& 194.7 &27.9& 0.024 & 840&  0.68& 2.7 &0.91  &2.55 & 357 & 11.7 \\
 2 &A1367&   90 & 245& 176.2 &20.0& 0.023 & 685&  0.51& 2.0 &0.46  &1.15 & 400 & 8.5 \\
 3 &A1185&  607 & 211& 167.6 &28.4& 0.034 & 610&  0.61& 2.2 &0.47  &1.15 & 409 & 6.4 \\
\hline                                        
\label{tab:cl}  
\end{tabular}\\
\tablefoot{                                                                                 
Columns are as follows:
(1): Order number of the cluster;
(2): Abell ID number of the cluster;
(3): ID of the cluster from \citet{2014A&A...566A...1T};
(4): Number of galaxies in the cluster, $N_{\mathrm{gal}}$;
(5)--(6): cluster centre right ascension and declination (in degrees);
(7): redshift of the cluster;
(8): line-of-sight velocity dispersion of cluster member galaxies, in \mbox{km\,s$^{-1}$};
(9): cluster gravitational radius $R_{\mathrm{g}}$ (in $h^{-1}$ Mpc);
(10): maximum radius of a cluster in the plane of the sky (in $h^{-1}$ Mpc);
(11): mass of the cluster assuming an NFW density profile, $M_{\mathrm{NFW}}$,
(in $10^{15}h^{-1}M_\odot$);
(12): cluster total luminosity (in $10^{12} h^{-2} L_{\sun}$); 
(13): mass-to-light ratio (in $h\,M_\odot$/$L_\odot$);
(14): luminosity-density field value at the location of the cluster, $D8$, 
in units of the mean density as described in the text.
}
\end{table*}

\section{Methods} 
\label{sect:met} 

\subsection{Spherical collapse model} 
\label{sect:sphshort} 
We applied the spherical collapse model to describe the evolution of galaxy clusters
as collapsing spherical shells. 
Evolution of galaxy clusters in these models
can be described using several important epochs with characteristic density contrasts. 
In an expanding $\mathrm{\Lambda}$CDM universe, all the densities and density contrasts 
depend on redshift. In the local Universe $(z = 0)$ characteristic
density contrasts for cosmological parameters applied in this paper are as follows. Density contrast for virialisation is $\Delta\rho_{vir} = 217$, 
turnaround density contrast $\Delta\rho_{ta} = 13.1$, and density contrast at present for the objects that will 
collapse in the future (future collapse)
is $\Delta\rho_{FC} = 8.73$. The density contrast $\Delta\rho_{ZG} = 5.41$ corresponds to so-called zero 
gravity (ZG), at which the radial peculiar velocity component of 
the test particle velocity equals the Hubble expansion
and the gravitational attraction of the system and its expansion are equal. 
We describe the spherical collapse model
and its characteristic epochs with their density contrasts 
calculations as a function of redshift and other parameters in detail in Sect.~\ref{sect:sph}.

Undoubtedly, applying the spherical collapse model is an approximation only as nearly all real 
galactic systems are non-spherical. However, as demonstrated by \citet{Korkidis:2020}, 
comparing results of N-body simulations with the spherical collapse model calculations, 
this model also characterises the evolution of non-spherical 
systems (galaxy clusters) rather well. Thus, this justifies approximation in our analysis. 

We defined the turnaround region around a cluster
as the region at which border the  ``observed'' density contrast is equal to the density 
contrast of turnaround, $\Delta\rho_{ta}$. 
Under the ``observed'' we mean the densities we calculated from the masses
around clusters based on the dynamical masses of clusters and groups.
Based on this definition, 
we found the radius and mass within the turnaround region, $R_\mathrm{ta}$ and $M_\mathrm{ta}$
\citep[see also ][about the masses of superclusters]{2025Univ...11..167E}. 
Also, we found the radius and mass around clusters within the virialisation region, $R_\mathrm{vir}$ and $M_\mathrm{vir}$ 
using density contrast $\Delta\rho_{vir}$, and within the future collapse region, 
$R_\mathrm{FC}$ and $M_\mathrm{FC}$
using density contrast $\Delta\rho_\mathrm{FC}$.
We used these data  to predict the future evolution of clusters and the whole superclusters. 

For a spherical volume, 
the ratio of the density to the mean density (overdensity),
$\Delta\rho = \rho/\rho_{\mathrm{m}}$, at different redshifts can be calculated as
(see Eq.~\ref{eq:den_red})
\begin{equation}
\Delta\rho=6.88\,\Omega_\mathrm{m0}^{-1}\left(\frac{M}{10^{15}h^{-1}M_\odot}\right)
        \left(\frac{R}{5h^{-1}\mathrm{Mpc}}\right)^{-3}\cdot (1+z)^{-3},
\label{eq:sph}
\end{equation}
 where $\Omega_{m0}$ is the matter density parameter at present.

\begin{figure}[ht]
\centering
\resizebox{0.44\textwidth}{!}{\includegraphics[angle=0]{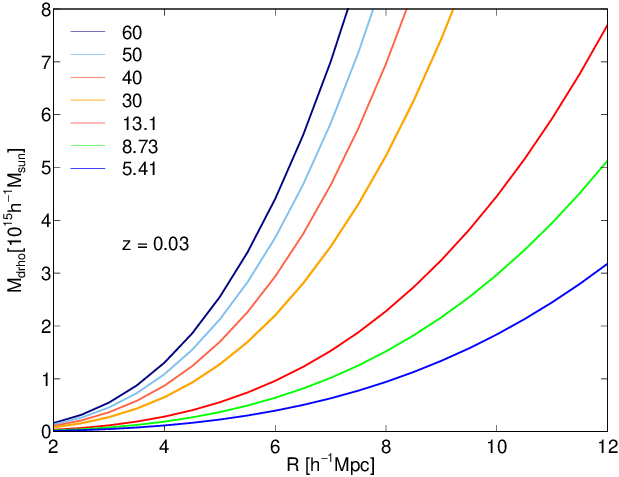}}
\caption{
Mass--radius relation from the spherical collapse model.
Different lines correspond to the masses for different density contrasts, $\Delta\rho$
(Eq.~\ref{eq:sph}). Lines $M_{60}$--$M_{30}$ correspond to the density contrasts, $\Delta\rho_{inf}$, 
at the borders of the spheres of influence as found for clusters. 
Lines $M_{13.1}$, $M_{8.73}$, and $M_{5.41}$
corresponds to the turnaround, future collapse, and zero gravity density contrasts. 
All lines are calculated for the redshift $z = 0.03$.
}
\label{fig:massradius}
\end{figure}

Using Eq.~(\ref{eq:sph}) we can calculate the relation between the mass and radius 
of an overdensity at various density contrasts and redshifts. 
For the mass within a sphere of influence, the density contrast in Eq.~(\ref{eq:sph}) 
corresponds to the density contrast, $\Delta\rho_{inf}$, determined in this study. 
In Fig.~\ref{fig:massradius} we show the mass--radius relation calculated
using the cosmological parameters used in our study.
In this figure, different lines show the mass versus radius of a spherical shell 
for a series of density contrasts (the turnaround, future collapse, and zero gravity),
and for the density contrasts at the borders of the spheres of influence, $\Delta\rho_{inf}$,
as shown in this study.

\subsection{Substructure of clusters} 
\label{sect:sub} 
We searched for possible substructure of rich clusters  in two ways.
First,  we applied the package {\emph{mclust}} for multidimensional normal mixture modelling,
 classification and clustering
\citep{fraley2006} from {\emph{R}}  statistical environment 
\citep{ig96}\footnote{http://www.r-project.org}.
{\emph{mclust}} package have been used to search for substructure in galaxy clusters in, for example,
\citet{2012A&A...540A.123E}.
This package studies a  finite mixture of distributions, 
in which each component is taken to correspond to a different subgroup of the cluster.
{\emph{mclust}}  finds for each datapoint the probability to belong to a component. 
The mean uncertainty for the full sample  is a statistical estimate of the reliability
of the results. As an input for {\emph{mclust},} we used 
the sky coordinates and line-of-sight velocities 
(calculated from their redshifts) of the cluster member galaxies. 
The best solution for the components 
was chosen using the Bayesian information criterion (BIC). The algorithm finds components, their 
membership and probabilities for galaxies to belong to a component simultaneously.
\citet{2012A&A...540A.123E} tested how 
the errors in line-of-sight velocities of galaxies affect 
the reliability of the results of {\emph{mclust}}. They applied test in which they
shifted randomly the peculiar velocities of galaxies; these random shifts were
chosen from a Gaussian distribution with the dispersion equal to
the velocity dispersion of galaxies in a cluster. Such tests were performed
1000 times.  The number
of the components found by {\emph{mclust}} remained unchanged,
demonstrating that the results are robust against such errors.
Second, we compared the substructure obtained in this way with partner clusters 
from WH24m  catalogue \citep{2024MNRAS.532.1849W}.

\begin{figure*}
\centering
\resizebox{0.90\textwidth}{!}{\includegraphics[angle=0]{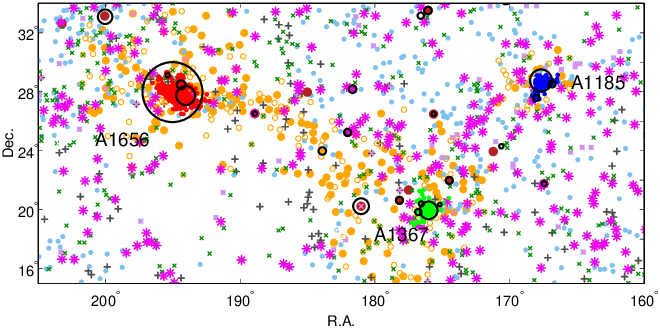}}\\
\resizebox{0.42\textwidth}{!}{\includegraphics[angle=0]{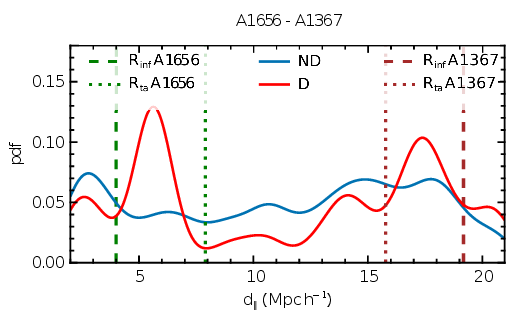}}
\resizebox{0.42\textwidth}{!}{\includegraphics[angle=0]{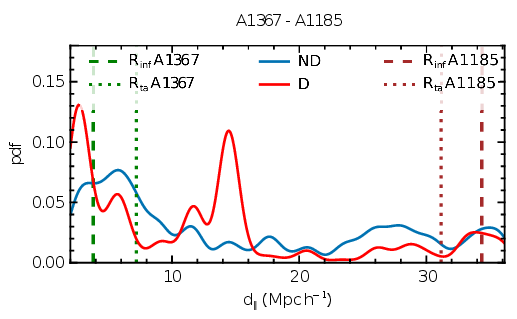}}\\
\caption{Upper panel: Sky distribution of groups (filled circles and crosses) and single galaxies
(empty circles) in the region of the Coma and the Leo superclusters.
Colours of symbols show groups in different global luminosity-density regions
(orange $D8 \geq 5$, blue $5 > D8 \geq 1.5$, 
and grey crosses $D8 < 1.5$). To avoid strong projections, we only plot single galaxies
in superclusters ($D8 \geq 5$).
Filament member galaxies in groups are shown with magenta x-s and single galaxies with green x-s.
Violet stars show members of long filaments with length $F_{\mathrm{len}}^{l} \geq 5$~\Mpc.
In groups we show the location of the brightest group galaxies only.
Dark red circles: $L_{gr} \geq 15\times10^{10} h^{-2} L_{\sun}$. 
Member galaxies of clusters A1656, A1367, and A1185 are shown in red, green, and blue.
Black circles mark the location of  groups from WH24 catalogue, circle sizes are proportional to group
richness. 
Lower panels: Number density distribution of galaxies in groups and single galaxies 
between clusters A1656 and A1367 (left panel), 
and A1367 and A1185 (right panel). 
Here, $d_\parallel$ denotes the coordinate on a straight line between the clusters, 
and red and blue lines show the linear density and linear number density distributions, respectively. 
The cross-section radius, where the densities were evaluated, 
was $8\,{\rm Mpc\,h^{-1}}$. The vertical dashed and dotted lines show influence and turnaround 
radii locations away from their respective end-point clusters. 
}
\label{fig:skygr15wh24}
\end{figure*}

\subsection{Projected phase space diagram and various infall regions} 
\label{sect:pps} 

In the PPS diagram we plotted line-of-sight velocities of galaxies 
with respect to the cluster mean velocity versus the projected clustercentric distance of galaxies.
Simulations show that in this diagram, galaxies at small projected clustercentric 
distance form an early infall population with 
infall times $\tau_{\mathrm{early}} > 1$~Gyr. The early infall region approximately corresponds to the
virialised part of the clusters. Galaxies 
at large projected clustercentric distance
form late infall populations with $\tau_{\mathrm{late}} < 1$~Gyr
\citep{2013MNRAS.431.2307O, 2014ApJ...796...65M, 2015ApJ...806..101H, 2024MNRAS.535.1348C}. 
Borders of these infall regions can be approximately 
found as follows \citep{2013MNRAS.431.2307O}:
\begin{equation}\label{eq:infalltime}
\left\lvert \frac{v-v_\text{mean}}{\sigma_\text{cl}} \right\lvert= 
-\frac{4}{3}\frac{D_\text{c}}{R_\text{vir}}+2 , 
\end{equation}
where $v$ are the velocities of the galaxies, $\sigma_\text{cl}$ is the velocity dispersion of galaxies in
the cluster,
$D_\text{c}$ is the projected clustercentric distance of galaxies, and 
$R_{\mathrm{vir}}$ is the cluster virial radius. 
Galaxies at small clustercentric distances may have been infallen at redshifts
$z \geq 1$,
as suggested by ages of stellar populations of the brightest cluster galaxies
\citep{2022A&A...668A..69E}. 

Gravitational radius of clusters, $R_{\mathrm{g}}$, is taken from \citet{2014A&A...566A...1T}. 
This is the radius used in the scalar virial theorem \citep[][Eq.~4.2.49a]{2008gady.book.....B}  
and it should not be confused with the cosmological virial radius 
\citep[][Eq.~9.66]{2008gady.book.....B}.
We used
the projected phase space diagram 
to determine regions of late infall of galaxies with characteristic radius $R_{\mathrm{cl}}$. 
In the PPS diagram  $R_{\mathrm{cl}}$ is marked by a small minimum in the
clustercentric distance distributions of galaxies 
\citep{2020A&A...641A.172E, 2021A&A...649A..51E}. 
Usually $R_{\mathrm{cl}}$ is called the splashback radius---radius,
at which the density profile of a cluster changes, and 
the slope of the correlation function
of galaxies also changes, which is a signature of 
the crossover from galaxy correlations within a cluster to 
correlations in small groups in filaments between clusters \citep{1991MNRAS.252..261E, 
2015ApJ...810...36M, 2015ApJ...806..101H, 2017ApJ...843..128R}.
At this radius outer (infalling) components and substructures
reach the main component (main cluster) and galaxy orbits change \citep{2015ApJ...810...36M}.
The value of this radius and how it is related to the other radii of the cluster
depends on the structure and dynamical history of the cluster, and different probes may give different values for this radius
\citep{2015ApJ...810...36M, 2024A&A...689A..19L}.
 As we did not have data on galaxy orbits in our study, we called this radius as cluster radius $R_{\mathrm{cl}}$.
This radius approximately borders the late infall region of the cluster.
We applied Eq.~\ref{eq:infalltime} with $R_{\mathrm{cl}}$ to show a late infall time region in the PPS diagram.
Another radius which characterises clusters is their maximum size on the sky, $R_{\mathrm{max}}$,
provided in \citet{2014A&A...566A...1T}.

The border of the sphere of influence of groups is marked by another small minimum in the galaxy 
distribution around clusters in the PPS diagram. It defines the 
radius of the spheres of influence, $R_\mathrm{inf}$ \citep{2020A&A...641A.172E, 2021A&A...649A..51E}. 
Within the sphere of influence,  galaxies and groups are falling into clusters.
This is what \citet{2021MNRAS.503.4250F} proposed to call the depletion radius.
We calculated the distribution of mass around clusters as the sum of masses of the main
cluster and groups in these regions and used it to calculate the mass within the region of 
influence, $M_\mathrm{inf}$, and corresponding density contrast, $\Delta\rho_\mathrm{inf}$.
With spherical collapse model and $\Delta\rho_\mathrm{inf}$ we found 
 at which redshifts regions of influence  were at turnaround, and when in the future these regions will virialise.

\subsection{Connectivity of clusters} 
\label{sect:con}

The 3D connectivity of clusters is defined as the number of galaxy filaments, $C = N_{\mathrm{fil}}$,
connected to a cluster \citep{2000PhRvL..85.5515C, 2018MNRAS.479..973C}.
However, detecting filaments around clusters using observational data may be 
complicated due to the sometimes messy structures surrounding clusters
(e.g. extended substructures) and infalling galaxies and groups.
Elongated groups and also elongated substructures of clusters 
may be (mis)classified as short filaments. In order to determine the connectivity
of clusters, we used Bisous filaments, as described above, and excluded very short filaments 
with a length of $F_{\mathrm{len}} < 3$~\Mpc, as they are less reliable.
We note that galaxies are considered filament members 
if their distance to the nearest filament axis is $D_{fil} \leq 0.5$~\Mpc.
We defined long filaments 
as filaments with a length of $F_{\mathrm{len}}^{l} \geq 5$~\Mpc\ 
and short filaments as those with a length of $3 < F_{\mathrm{len}}^{s} < 5$~\Mpc.
We found the total number of filaments, $N_{\mathrm{fil}}^{all}$, 
and the number of long filaments, $N_{\mathrm{fil}}^{l}$,
for each cluster within the region of influence of a cluster. 
This approach was used to determine the connectivity
of the richest clusters in the A2142 and the Corona Borealis superclusters
in \citet{2020A&A...641A.172E} and \citet{2021A&A...649A..51E}.
We compared our results 
with those obtained by \citet{2020A&A...634A..30M}, who used 
the Discrete Persistent Structure Extractor 
(DisPerSe)
to extract filaments in the Coma supercluster region, and with those by \citet{2024NatAs...8..377H},
who studied intracluster filaments in the Coma cluster.

\section{Results }
\label{sect:results}  

We present the results of our analysis in this section. 
We start with the overall analysis of the luminosity-density field and 
galaxy  and group content of 
the region under study. Then we
analyse the substructure and connectivity of clusters. Next, we determine 
different infall regions of clusters with their characteristic 
radii and calculate the density contrasts around clusters based on the mass distribution. 
Then we analyse the galaxy content in clusters, in their regions of influence,
and in superclusters and low-density regions around clusters,
and we analyse the relation between masses of clusters and masses embedded in their 
regions of influence and turnaround.

\subsection{Structure of the Coma and Leo superclusters}
\label{sect:glob}  

We used the luminosity-density field to determine supercluster borders and to separate galaxies, 
groups, and clusters in superclusters and in surrounding low-density regions. 
Figure~\ref{fig:skygr15wh24} (upper panel) shows the sky distribution of groups in our sample
in a sky area covering the Coma and the Leo superclusters. 
 We see that the luminous groups are mostly located
in supercluster regions with  $D8 > 5.0$, together with very poor groups and single galaxies. 
This figure 
shows also, that while in the environment of A1656 and A1185 
galaxy filaments are dominating, the environment of A1367 is rich in galaxy groups
\citep[see also ][who discussed groups surrounding this cluster]{2020MNRAS.497..466S}. 
Cluster environments were also discussed in the early study of the Coma supercluster \citep{1978ApJ...222..784G}.
There is a minimum in the group and galaxy distribution around each 
cluster at distances $\approx 8 - 10$~\Mpc. We shall discuss this in Sect.~\ref{sect:discussion}.

Figure~\ref{fig:skygr15wh24} shows that there is a quite clear filament connecting 
clusters A1656 and A1367. The clusters  A1367 and A1185 are separated by low-density environment
(blue symbols, and the lack of orange symbols between these clusters, Fig.~\ref{fig:skygr15wh24}). 
In the lower panels of  Fig.~\ref{fig:skygr15wh24}
we show the linear density of galaxies versus their distances 
along the straight line connecting the clusters
from A1656 to A1367 (left panel), and from A1367 to A1185 (right panel). 
We can notice that along the filaments, there are minima in the grouped galaxy distributions at distances 
$\approx 4$~\Mpc\ and  $\approx 8$~\Mpc\  from each cluster.

\subsection{Substructure and connectivity of clusters}
\label{sect:subcon}  

\begin{figure*}[ht]
\centering
\resizebox{0.33\textwidth}{!}{\includegraphics[angle=0]{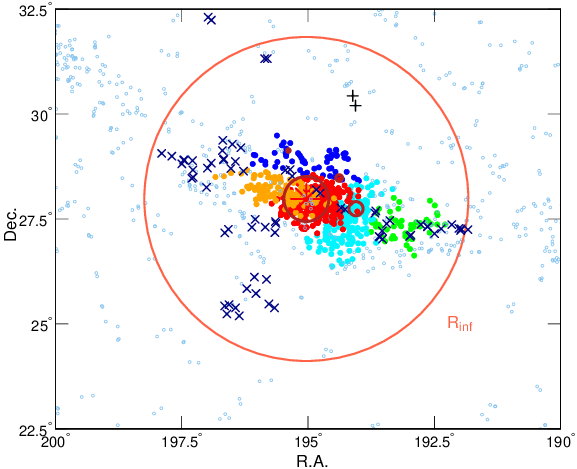}}
\resizebox{0.33\textwidth}{!}{\includegraphics[angle=0]{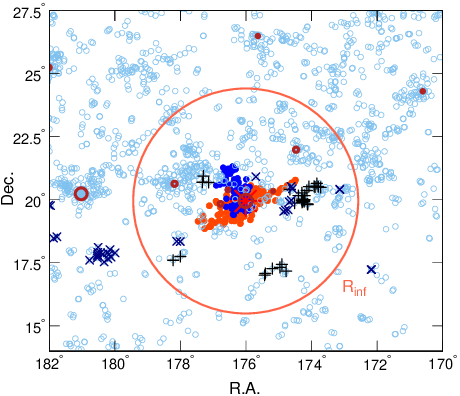}}
\resizebox{0.33\textwidth}{!}{\includegraphics[angle=0]{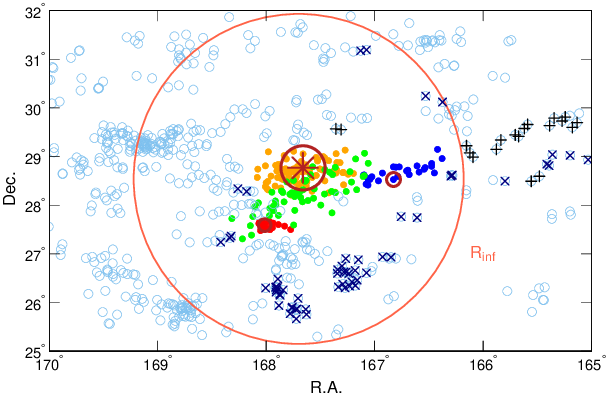}}\\
\vspace{0.5cm}
\resizebox{0.33\textwidth}{!}{\includegraphics[angle=0]{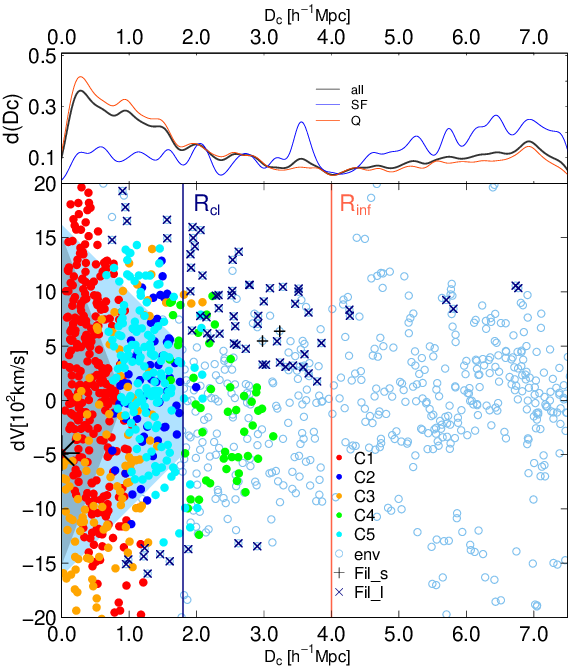}}
\resizebox{0.33\textwidth}{!}{\includegraphics[angle=0]{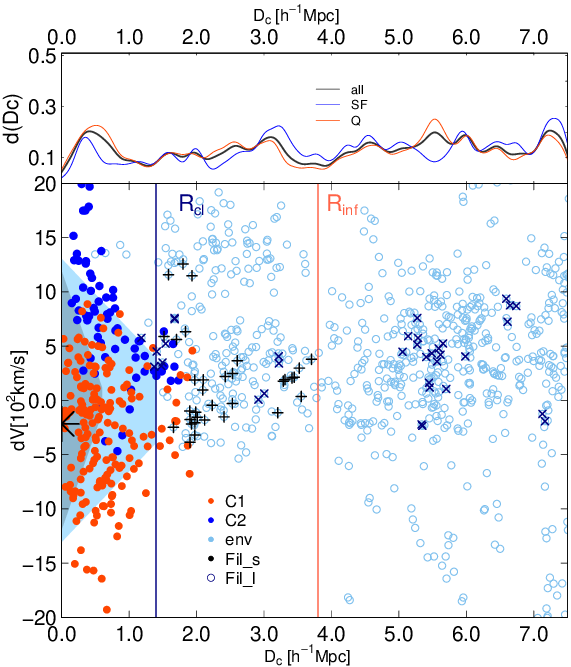}}
\resizebox{0.33\textwidth}{!}{\includegraphics[angle=0]{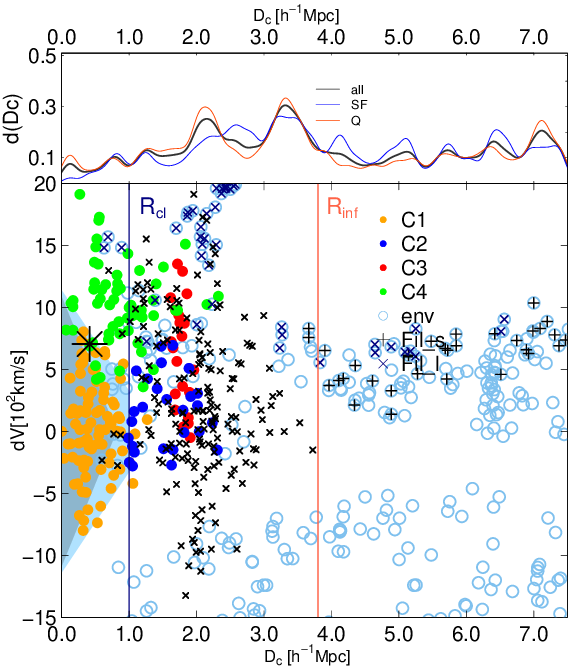}}\\
\vspace{0.5cm}
\resizebox{0.33\textwidth}{!}{\includegraphics[angle=0]{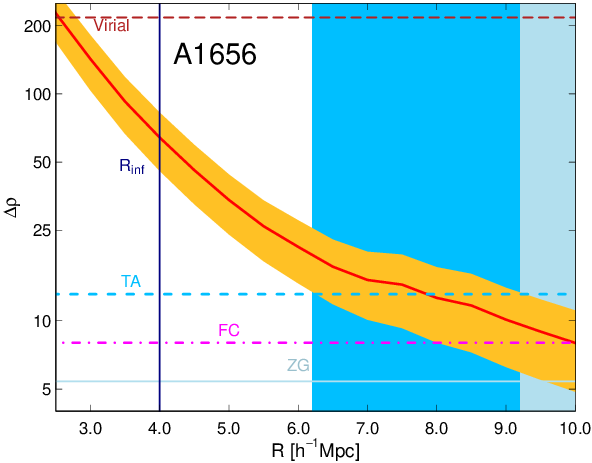}}
\resizebox{0.33\textwidth}{!}{\includegraphics[angle=0]{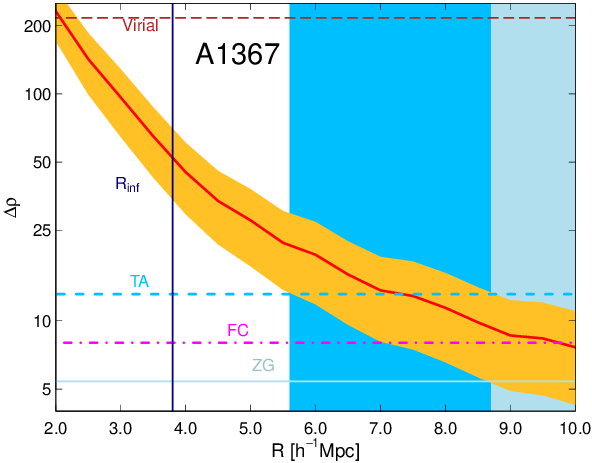}}
\resizebox{0.33\textwidth}{!}{\includegraphics[angle=0]{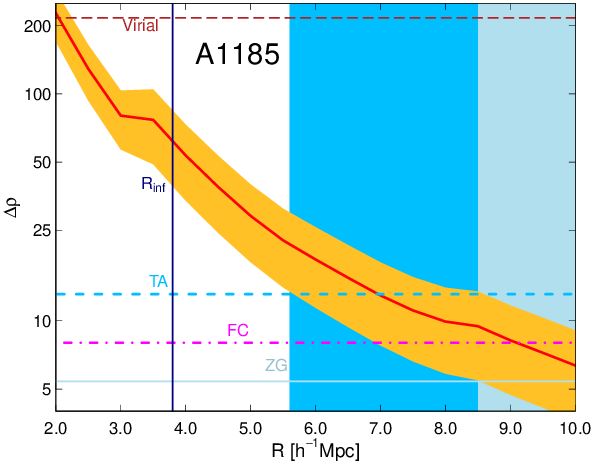}}
\caption{Upper row: 
Distribution of galaxies in clusters and in their environment in the plane of the sky.
Left: A1656. Middle: A1367. Right: A1185.
Symbols of different colours show galaxies in different components of the cluster:
the redder the colour, the higher the median values of the $D_n(4000)$ index of galaxies are
in a component.
Black crosses show galaxies of short filaments ($F_{\mathrm{len}}^{s} \leq 5$~\Mpc),
and dark blue crosses show galaxies in long filaments ($F_{\mathrm{len}}^{l} > 5$~\Mpc).
Dark red circles show clusters from WH24m catalogue. Circle sizes are proportional to the
number of galaxies in a cluster. Orange circles show the regions
of influence, with $R_{inf}$.
Middle row:
Distribution of the clustercentric distances, $D_c$ 
(upper panels), and the PPS diagram (lower panels).
Symbols are as in the upper panels.
Vertical lines show the cluster radius, $R_\mathrm{cl}$, 
and the radius of the sphere of influence, $R_\mathrm{inf}$.
In the lower panel, the dark blue region is the early infall region, 
and light blue is the late infall region.
Lower row:
Density contrast, $\delta \rho$, versus clustercentric distance. 
Vertical line shows the radius of the sphere of influence, $R_\mathrm{inf}$, and
horizontal lines mark the characteristic density contrasts, as shown in the Figure. The blue area shows turnaround  region, and the light blue area shows future collapse region.
}
\label{fig:clskyppsdrho}
\end{figure*}

To analyse the substructure of clusters, 
we found components of each cluster, using SDSS data and applying 
{\emph{mclust}}. The number of components, $N_\mathrm{comp}$, and their properties are given in Table~\ref{tab:sub}.
Independently, we found from WH24m  partner cluster systems 
(Table~\ref{tab:wh24cl}). 
The number of components, $N_\mathrm{comp}$, and the number of partners using WH24 and 
WH24m data, $N_\mathrm{part}$,  as well as other main parameters of the three main clusters 
are given in Table~\ref{tab:clresults}. 

In Fig.~\ref{fig:clskyppsdrho}, for each cluster, we show the 
distribution of galaxies in the cluster and around
it in the sky plane (upper row), as well as the 
PPS diagram (middle row), and the density contrast $\delta\rho$,
calculated from the distribution of mass around clusters,
versus clustercentric distance $R$ (lower row). In the sky plane plot in upper panels, we show the border of the sphere of 
influence and mark partner clusters from the WH24m catalogue.

Table~\ref{tab:clresults} and Fig.~\ref{fig:clskyppsdrho} show that all clusters in our study have 
substructure. The Coma cluster, A1656, consists of two main components 
(components 1 and 3 in Table~\ref{tab:clresults}) and three smaller components 
in the outer parts of the cluster (Fig.~\ref{fig:clskyppsdrho})
\citep[see also][where two main components were discussed]{1997ApL&C..36...97D}. 
Component 4 may be an endpoint of an infalling filament \citep{2024NatAs...8..377H}. Interestingly, 
A1656
is often considered as a typical relaxed cluster \citep{2006PASP..118..517B, 2020MNRAS.497..466S}. 
In contrast and in agreement with our results, other studies found that the Coma cluster 
represents a typical cluster with substructure \citep[see][for an early review]{1998ucb..proc....1B}. 
In agreement with this, \citet{2023RMxAA..59..345C} showed that A1656 has a significant substructure in which massive component
(central cluster) is accreting smaller groups.
A recent analysis of intracluster light in the A1656 suggests that even the main component 
of this cluster has subclumps, and the brightest galaxies in it are in the 
process of merging \citep{2025A&A...694A.216J}.

The cluster A1367 is known as a dynamically active cluster with two components and a merger shock 
\citep[see][and references therein]{2022A&A...665A.155D, 2019MNRAS.486L..36G}. Our 
analysis revealed 
these components, and  a short filament in the direction of merger shock 
described by \citet{2019MNRAS.486L..36G}. 

In cluster A1185, usually the galaxy NGC~3550 is considered as the main (brightest) galaxy 
\citep{2011A&A...528A.115W}. However, in \citet{2014A&A...566A...1T}, 
this is the second brightest galaxy in this cluster.
The brightest galaxy and the second brightest galaxy both lie in the main component of the cluster
and are offset with its X-ray centre \citep{2011A&A...528A.115W}.
Main component of A1185 hosts also interacting galaxy NGC~3561 (Arp~105), a signature of 
dynamical  activity  \citep{2022hst..prop17212W}. 
One component, identified with the WH24 cluster J110718.2+283140, maybe a terminal part of a 
long filament. Another WH24 cluster, J111203.3+273523, may represent a group, falling into
cluster along filament which points towards A1367. 

Our results of substructure analysis are in a good agreement with 
partner systems from the WH24m catalogue. Main clusters in the WH24m partner systems can be 
identified with the richest components among substructures, and partners with poor components. 
There are two exceptions.  
WH24 cluster $J115242.6+2037534$, one of the partner clusters of the 
cluster A1367, corresponds in our catalogue to a short filament 
within it's sphere of influence. Poor cluster $J114612.2+202330$
is not among A1367 partners, probably due to a redshift difference
and strict definition of partner clusters in WH24m.
When we compared the masses of clusters, the sum of the masses of partner cluster systems 
from WH24m agreed well with the cluster masses from SDSS data.
However, our analysis did not detect fine details in each component of clusters,
as, for example, in \citet{2025A&A...694A.216J} for the Coma cluster.
This comparison shows that, in fact, in different studies the same structures 
have been found, but, due to the different definitions, these systems may have been identified differently
(short filament versus partner cluster versus cluster component). 
Also, the use of {\emph{mclust}} have its limitations. It assumes Gaussianity, and it performs
better with a large number of galaxies in a cluster \citep{2013MNRAS.434..784R}.
However, the good agreement between different substructure and partner cluster analysis shows that
{\emph{mclust}} gives reliable results.

{\emph{Connectivity of clusters}}. 
To determine the connectivity of clusters, we 
first analysed the PPS diagram of clusters (Fig.~\ref{fig:clskyppsdrho}, middle row) in order to find
the radius of the sphere of influence of clusters, $R_\mathrm{inf}$.
The border of the sphere of influence is marked by a small minimum in the galaxy 
distribution around clusters in the PPS diagrams. 
The values of $R_\mathrm{inf}$ are given in Table~\ref{tab:clresults}. 
The radius of influence for A1656, $R_{inf}^{A1656} \approx 4$\Mpc\ agrees well 
with the radius which was determined 
by \citet{2025arXiv250404135B} as a distance from the cluster centre, at which
the velocities of galaxies change. 
Also other clusters have approximately the same radii of influence, $R_\mathrm{inf} \approx 3.8$\Mpc.

The median value of the ratio of radii $R_\mathrm{cl}$ and $R_\mathrm{max}$ 
in our study is $R_{\mathrm{cl}}/R_{\mathrm{max}} \approx 0.71$.
\citet{2022A&A...664A.198G} showed that the
parameter which characterises the shape of clusters
seems to change at relative radius $r_\mathrm{eff} \approx 0.75$, which approximately corresponds to our 
ratio $R_{\mathrm{cl}}/R_{\mathrm{max}}$.
As $R_{\mathrm{cl}}$ is close to the radius at which galaxy orbits change,
the change in the shape parameter in \citet{2022A&A...664A.198G} may be related to that. 

Next, we identified galaxy filaments within $R_\mathrm{inf}$,
which defines the 3D connectivity of clusters. 
We provide in Table~\ref{tab:clresults} the number of all filaments $N_\mathrm{fil}^{all}$, 
and the number of long filaments, $N_{\mathrm{fil}}^{long}$, connected to a given cluster.
For comparison with previous work, we show in Table~\ref{tab:clresults} 
also data on the characteristic radii, masses, and connectivity for the 
rich clusters in the Corona Borealis and in the A2142 superclusters.
Data  on these clusters
(A2065, A2061, A2089, Gr2064 in the Corona Borealis supercluster, and A2142
in the SCl~A2142) in Table~\ref{tab:clresults} are from \citet{2020A&A...641A.172E} and \citet{2021A&A...649A..51E}. 
We see that clusters in the Coma and Leo superclusters 
have lower connectivity than clusters in the Corona Borealis supercluster.
Especially high is the connectivity of the cluster of the highest mass, 
A2065. 

In this study we considered galaxies as filament members if
their distance to the nearest filament axis $D_{fil} \leq 0.5$~\Mpc.
Typical width of filaments is approximately $1$~\Mpc, as
also found from IllustrisTNG simulations for the present epoch 
by \citet{2025ApJ...989..187Y} ($1 - 1.5$~\Mpc\ in their study where they use 
DisPerSE filament finder
to identify filaments).
\citet{2020A&A...641A.172E} analyse how the use of different distances to the nearest
filament axis $D_{fil}$ affects the detection
of filaments and, consequently, the connectivity of clusters. 
The use of large $D_{fil}$ may merge close filaments near clusters
(as in the case of filaments at east side of the Coma cluster), decreasing the value
of connectivity of clusters. Also, the use of larger  $D_{fil}$ may make filaments wider and sparser,
or even create false filaments from sparse distribution of galaxies and groups. 
The exact change in filament properties depends on the configuration of galaxies near clusters. 
\citet{2020A&A...641A.172E} also analyse how the use of larger $D_{fil}$ may affect the connectivity of 
groups outside clusters, in low-density environment.
We refer to \citet{2020A&A...641A.172E}, Sect. 4 for the details of this analysis.

\citet{2020A&A...634A..30M} use the DisPerSE filament finder
to detect 
long filaments around A1656. In this study three long filaments are detected. 
This is less than in our 
study; the difference is related to the different methods to define filaments.
The structure at approximately $R.A. \approx 196^{\degr}$ 
and $Dec. \approx 26^{\degr}$ in Fig.~\ref{fig:clskyppsdrho} (upper left panel)
is identified as a filament in our study, but it is not among filaments in 
\citet{2020A&A...634A..30M} (see their Fig.2).

We note that components  in A1656 agree well with the positions of
intracluster filaments found by \citet{2024NatAs...8..377H} in A1656, based on weak lensing
analysis of Hyper Suprime-Cam imaging data (their N and W filaments). One short filament in our study
coincides with their SE filament.
Moreover, \citet{2017ApJ...845...24K} reported a discovery of a virial
shock region around A1656 with radius of about $5$~Mpc. The borders of this
shock region approximately agree with the borders of the region of influence around this cluster.
\citet{2020MNRAS.497.3204M} detected an extended X-ray emission in the direction of the filament from A1656
 towards A1367, seen also in Fig.~\ref{fig:clskyppsdrho}.
Another shock is related to components 4 and 5. Filament end W
 was also reported by \citet{2025ApJ...978L..47C}.
 
Two long filaments connected to the cluster A1367 have been noted earlier by 
\citet{2004A&A...425..429C} and \citet{2000ApJ...543L..27W}. \citet{2004A&A...425..429C} described
A1367 as dynamically young cluster
of two components, as also found in our analysis.

\begin{table*}[ht]
\caption{Properties of clusters in superclusters.}
\begin{tabular}{rrrrrrrrrrrrrr} 
\hline\hline  
(1)&(2)&(3)&(4)&(5)& (6)&(7)&(8)&(9)&(10)&(11)&(12)&(13)&(14)\\      
\hline 
No. & Abell ID& $N_{\mathrm{comp}}$ &$N_{\mathrm{part}}$&  $R_{cl}$ &  $R_\mathrm{inf}$ & $M_\mathrm{inf}$ & $\Delta\rho_\mathrm{inf}$
&  $R_{ta}$ & $M_{ta}$ &  $R_{FC}$ & $M_{FC}$ &$N_{\mathrm{fil}}^{all}$ & $N_{\mathrm{fil}}^{long}$  \\
\hline       
\multicolumn{5}{c}{Coma and Leo superclusters} &&&& \\
\hline                
 1 &A1656&  5 & 4 &1.8 & 4.0 & 1.3$\pm$0.4 & 60$\pm10$& 7.9 & 2.0$\pm$0.7  & 10.0 & 2.5$\pm$0.8 & 6 & 5  \\
 2 &A1367&  2 & 4 &1.4 & 3.8 & 0.8$\pm$0.4 & 50$\pm10$& 7.2 & 1.5$\pm$0.5  & 9.8& 2.0$\pm$0.8 & 6 & 2  \\
 3 &A1185&  4 & 3 &1.0 & 3.8 & 1.0 $\pm$ 0.3 & 60{\raisebox{0.5ex}{\tiny$^{+10}_{-15}$}}& 7.0 & 1.4$\pm$0.5 & 8.0 & 1.8$\pm$0.7 & 5 & 3 \\
\hline                
\hline                
\multicolumn{5}{c}{Corona Borealis and A2142 superclusters} &&&& \\
\hline                
 4 &A2065&  4 &     &2.5 & 6.0 & 2.6  & 30 & 9.5  & 3.7 &11.9 & 4.3& 9 & 4  \\
 5 &A2061&  2 &     &1.4 & 4.0 & 0.9  & 40 & 6.4  & 1.2 &8.6 & 1.8 & 8 & 5 \\
 6 &A2089&  1 &     &1.3 & 3.0 & 0.5  & 45 & 5.6  & 0.8 &6.9 & 0.9 & 2 & 1  \\
 7 &Gr2064& 3 &     &1.4 & 4.0 & 0.7  & 30 & 4.9  & 0.7 &8.5 & 1.6&  6 & 3 \\
\hline                                
 8 &A2142&  5 &     &1.8 & 5.0 & 1.8  & 30 & 8.0  & 2.3 & 9 & 2.4 &6--7& 6--7 \\
\hline                                        
\label{tab:clresults}  
\end{tabular}\\
\tablefoot{                                                                                 
Columns are as follows:
(1): Order number of the cluster;
(2): Abell ID number of the cluster;
(3): Number of substructures;
(4): Number of clusters and in cluster partner systems in WH24 and WH24m catalogues;
(5): Cluster radius (see text for definition), $R_{cl}$, in \Mpc;
(6): Radius of the sphere of influence, $R_\mathrm{inf}$, in \Mpc;
(7): Mass embedded in the sphere of influence, $M_{inf}$, in $10^{15}h^{-1}M_\odot$;
(8): The density contrast at the border of the sphere of influence, $\Delta\rho_\mathrm{inf}$;
(9): Radius of the turnaround region, $R_{ta}$, in \Mpc;
(10): Mass embedded within the turnaround region, $M_{ta}$, in $10^{15}h^{-1}M_\odot$;
(11): Radius of the future collapse region, $R_{FC}$, in \Mpc;
(12): Mass embedded within the future collapse region, $M_{FC}$, in $10^{15}h^{-1}M_\odot$;
(13--14): The number of all filaments and long filaments filaments connected to a cluster,
$N_{\mathrm{fil}}^{all}$, 
and $N_{\mathrm{fil}}^{long}$ with length $F_{\mathrm{len}}^{l} \geq 5$~\Mpc.
}
\end{table*}

\subsection{Density contrasts, characteristic radii, and masses }
\label{sect:gal}  

In Fig.~\ref{fig:clskyppsdrho} (lower panels) we present the density contrast $\Delta\rho$
for each cluster in our study, calculated using the distribution of mass around clusters.
In this figure the 
blue area shows the turnaround region, and light blue area corresponds to the future
collapse region.
From Fig~\ref{fig:clskyppsdrho} we can find the characteristic density contrast at the 
borders of the regions of influence and the mass embedded within these regions, $\Delta\rho_{inf}$, and
$M_{inf}$.
Using Fig.~\ref{fig:clskyppsdrho} we also found radii and embedded masses for the characteristic
epochs of the spherical collapse model, namely, turnaround, future collapse, and 
zero gravity (with corresponding radii and masses $R_\mathrm{ta}$, $R_\mathrm{FC}$, $R_\mathrm{zg}$,
and  $M_\mathrm{ta}$, $M_\mathrm{FC}$, $M_\mathrm{zg}$).  
The values of these radii, masses, and density contrasts (for the spheres of influence)
are given in Table~\ref{tab:clresults}.

For A1656, characteristic radii can be compared with those presented in Fig. 2 by \citet{2015A&A...577A.144T},
which presents the so-called $\Lambda$ significance diagram -- another way to
show the relation between the mass (density contrast) and the radius of an object.
For the A1656 cluster  \citet{2015A&A...577A.144T} present several regions. 
At small radius ($R \approx 1.4$)  the figure shows the location of the A1656 cluster itself,
larger radius ($R \approx 4.8$) corresponds to the region
of influence around A1656, in a good agreement with our estimate. 
The largest radius for A1656 is $R \approx  14$ --
approximately the radius of the zero gravity region around the cluster.

\subsection{Galaxy content}
\label{sect:galcont}  

\begin{figure*}[ht]
\centering
\resizebox{0.48\textwidth}{!}{\includegraphics[angle=0]{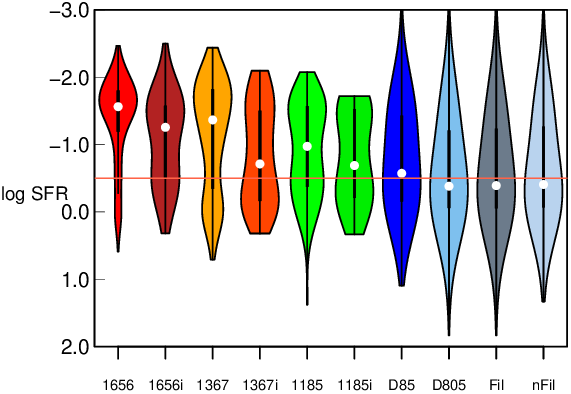}}
\resizebox{0.48\textwidth}{!}{\includegraphics[angle=0]{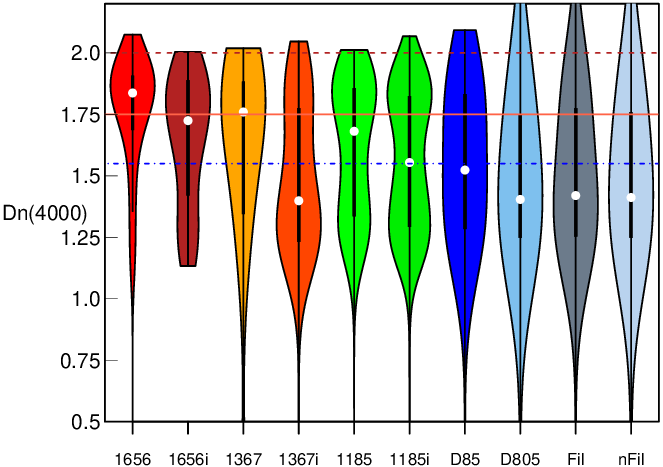}}
\caption{
Star formation rate $log SFR$ (left) and $D_{n}(4000)$ index for galaxies in
clusters, in the regions of influence (i) of clusters,
in the high-density regions ($D85$, excluding clusters; global luminosity-density
$D8 \geq 5$), and in the low-density regions ($D805$, $D8 < 5$ in the units of mean
luminosity-density) of our sample
and filament members (with distance from the nearest filament axis $D_{fil} <= 0.5$~\Mpc)
and non-member galaxies ($D_{fil} > 0.5$~\Mpc). Different samples are 
marked in the figure. We only considered filaments with length $L_{fil} \geq 3$~\Mpc\ as
more reliable.
The horizontal line in the left panel corresponds to $log SFR = -0.5$,
and lines in the right panel correspond to $D_{n}(4000) = 1.55$, $1.75$, and $2.0$.
}
\label{fig:viogal}
\end{figure*}

Next, we analysed the star formation properties of galaxies in clusters and around them, 
using star formation rates $\log \mathrm{(SFR)}$ and $D_{n}(4000)$ index. 
Distribution of these parameters are given in Fig.~\ref{fig:viogal} separately for 
galaxies in clusters and in the regions of influence of clusters, in high-density regions (excluding clusters)
and in low-density regions, in filaments and outside of filaments. 
Galaxies are considered as filament members if their distance 
to the nearest filament axis $D_{fil} <= 0.5$~\Mpc, otherwise they are located outside filaments.
For these subsamples
 we show in Table~\ref{tab:clgal} the median values of $\log\mathrm{(SFR)}$ and 
$D_{n}(4000)$.
Galaxy content of components in clusters is given in Table~\ref{tab:sub}.  
Calculations presented in Table~\ref{tab:sub} have been made 
using absolute magnitude limit $M_r \leq -18.3$, in order to compare magnitude-limited samples.

Table~\ref{tab:clgal} and Fig.~\ref{fig:viogal} show, as expected, that clusters and superclusters, 
in general, contain a higher percentage of quenched galaxies with no active star 
formation than regions around them,  and low-density regions between superclusters. 
We tested the statistical significance of the difference in galaxy populations 
between high and low global density environments and between filament member galaxies 
and those not in filaments using the Kolmogorov-Smirnov (KS) test. We considered that if $p \leq 0.01$
then the differences between distributions are highly significant. 
We did not apply the  KS test when samples were too small, with fewer than 20 galaxies
\citep{2013MNRAS.434..784R, 2024A&A...681A..91E}.

\begin{figure}[ht]
\centering
\resizebox{0.44\textwidth}{!}{\includegraphics[angle=0]{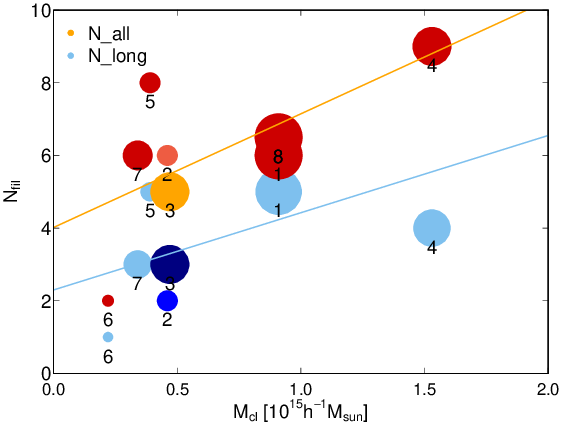}}
\caption{
Number of filaments connected to a cluster, $N_{\mathrm{fil}}$, versus  
the cluster mass, $M_{\mathrm{cl}}$.
Symbol sizes are proportional to the number of components in a cluster, $N_{\mathrm{comp}}$.
The numbers are order numbers of a cluster from Table~\ref{tab:clresults}.
Red symbols show the number of all filaments, and blue symbols the number of long
filaments with $F_{\mathrm{len}}^{l} \geq 5$~\Mpc. Light red and orange symbols and dark blue symbols correspond to clusters with higher values of
star-forming galaxies, A1367 and A1185 (clusters 2 and 3).
}
\label{fig:massfil}
\end{figure}

First, we compared galaxy populations in clusters A1656, A1367, and A1185, 
and found that these are different with
very high significance, with $p < 0.01$ between all cluster pairs. 
Galaxies with the oldest stellar populations, $D_{n}(4000) \geq 2.0$, reside in A1656 cluster,
and also in the environments of clusters.
In A1656 and A1367 the percentage of quenched galaxies in clusters is higher than in their 
regions of influence. In A1656, the main component and third component 
(orange symbols in Fig.~\ref{fig:clskyppsdrho}) embed the highest percentage of passive 
galaxies with old stellar populations among different components in this cluster. 
KS test tells that galaxy populations in 
A1656 and its region of influence are different with high significance level ($p < 0.01$).
The cluster A1367 and especially its region of influence, has 
a higher percentage of star-forming galaxies; populations are different with a 
high significance level, $p < 0.01$.
This was also noted by \citet{2007ApJ...658..929R}. 
In A1367 the main component has a higher percentage of passive galaxies with old stellar populations. 
The second component has a higher percentage of star-forming galaxies. 
The SFRs in the first and second component of A1367 are different with
high significance ($p < 0.01$), but the difference between the  
$D_{n}(4000)$ index values is not significant,  with $p < 0.11$. 
This is because of almost similar percentage of galaxies with very old stellar populations 
in these two components.

\begin{table*}[ht]
\caption{Galaxy content of clusters and
 areas around them.}
\begin{tabular}{rrrrrrrr} 
\hline\hline  
(1)&(2)&(3)&(4)&(5)&(6)&(7)&(8)\\      
\hline 
No. & Abell ID & $N_{\mathrm{gal}}^{cl}$ & $N_{\mathrm{gal}}^{inf} $ & $\log \mathrm{SFR}_{\mathrm{med}}^{cl}$ & $\log \mathrm{SFR}_{\mathrm{med}}^{inf} $ 
& $Dn4_{\mathrm{med}}^{cl} $ & $ Dn4_{\mathrm{med}}^{inf} $  \\
\hline                
 & & \multicolumn{2}{c}{Cl} &&&& \\
\hline                
 1 &A1656&  325 & 39 & -1.56 & -1.25 & 1.83 & 1.72 \\
 2 &A1367&  134 &279 & -1.37 & -0.65 & 1.76 & 1.40 \\
 3 &A1185&  144 &306 & -0.97 & -1.19 & 1.68 & 1.55 \\
\hline                
   &     &  $N_{\mathrm{gal}}^{D8>5}$ &$N_{\mathrm{gal}}^{D8<5} $ &  $\log \mathrm{SFR}_{\mathrm{med}}^{D8>5}$ &$\log \mathrm{SFR}_{\mathrm{med}}^{D8<5} $  &  $Dn4_{\mathrm{med}}^{D8>5}$ &$Dn4_{\mathrm{med}}^{D8<5} $  \\
\hline                
   &     &  904 & 8016 & -0.57 & -0.38 & 1.52 & 1.40 \\
\hline                
   &     &  $N_{\mathrm{gal}}^{Fil}$ &$N_{\mathrm{gal}}^{nFil} $&  
   $\log \mathrm{SFR}_{\mathrm{med}}^{Fil}$ &$\log \mathrm{SFR}_{\mathrm{med}}^{nFil} $  &  $Dn4_{\mathrm{med}}^{Fil}$ &$Dn4_{\mathrm{med}}^{nFil} $  \\
\hline                
   &     &  2771& 5331& -0.39 & -0.41 & 1.42 & 1.41 \\
\hline                                        
\label{tab:clgal}  
\end{tabular}\\
\tablefoot{                                                                                 
Columns are as follows:
(1): Order number of the cluster;
(2): Abell ID number of the cluster;
(3): Number galaxies with $M_r \leq -18.3$ with measured star formation rate and $D_{n}(4000)$ values in the cluster;
(4): Number galaxies with $M_r \leq -18.3$ with measured star formation rate and $D_{n}(4000)$ values in the region of influence of the cluster.
(5--8): Median values of star formation rate and $D_{n}(4000)$ index, $\log \mathrm{SFR}_{\mathrm{med}}$ and $Dn4_{\mathrm{med}}$ in clusters,
in the regions of influence of clusters, and the environment defined by global density field and filament membership.
}
\end{table*}

\begin{figure*}[ht]
\centering
\resizebox{0.44\textwidth}{!}{\includegraphics[angle=0]{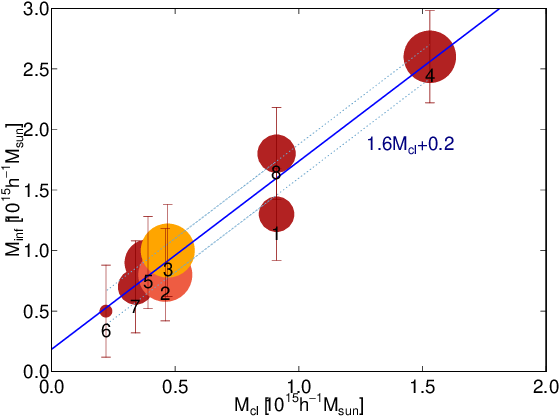}}
\resizebox{0.44\textwidth}{!}{\includegraphics[angle=0]{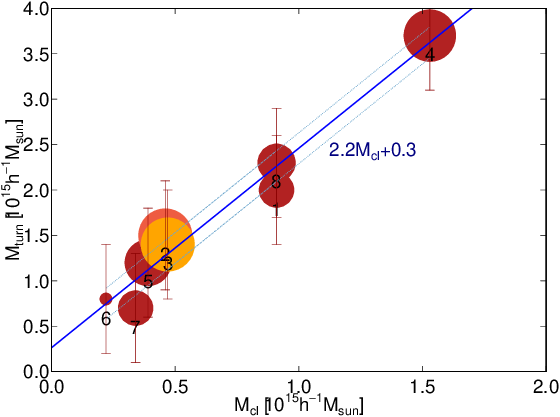}}
\caption{
Scaling relations between the cluster mass, $M_{\mathrm{cl}}$, and 
the mass embedded in the sphere of influence ($M_{\mathrm{inf}}$, left panel) and 
turnaround mass ($M_{\mathrm{ta}}$, right panel).
Symbol sizes are proportional to the number of all filaments connected to a cluster, $N_{\mathrm{fil}}$.
Dark red symbols show clusters with a high percentage of quiescent galaxies, 
and light red and orange symbols show clusters with a lower percentage of quiescent 
galaxies (light red - A1367, and orange - A1185).
Dashed lines show 1$\sigma$ errors of the linear fit, and error bars show mass errors (Table~\ref{tab:clresults}).
Numbers are order numbers of a cluster from Table~\ref{tab:clresults}.
}
\label{fig:massmass}
\end{figure*}

\begin{figure}[ht]
\centering
\resizebox{0.44\textwidth}{!}{\includegraphics[angle=0]{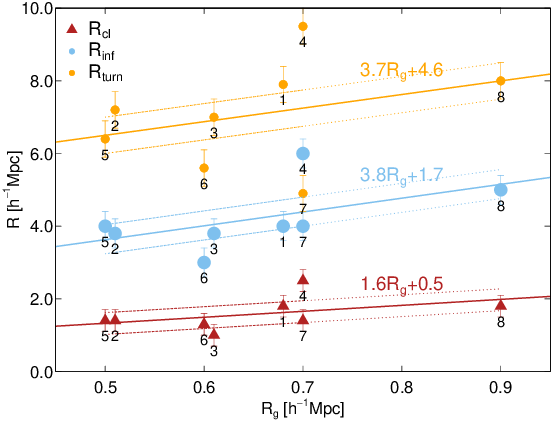}}
\caption{
Scaling relations between the cluster gravitational radii, $R_g$, and cluster radii, $R_{cl}$ (dark red); 
the radii of the regions of influence, 
$R_{inf}$ (blue); and the radii of the turnaround regions, $R_{ta}$ (orange).
Parameters of the linear scaling relations are given in the figure. 
Dashed lines show 1$\sigma$ errors of linear fit, and error bars show radii errors.
Numbers are order numbers of a cluster from Table~\ref{tab:clresults}.
}
\label{fig:rgrrr}
\end{figure}

For A1185 we see the opposite -- the cluster itself has a higher percentage of star-forming 
galaxies than its surrounding region. According to the SFR, the star formation properties 
of galaxies in the cluster and in it's sphere of influence are different with high 
significance ($p < 0.01$). However, the significance of differences is 
low if we consider the $D_{n}(4000)$ index, with $p < 0.3$. 
This mainly comes from the low percentage of quenched galaxies in its main component 
(Table~\ref{tab:sub}), and it
is related to a lower percentage of galaxies with very old stellar populations, 
$D_{n}(4000) \geq 1.75$ in this cluster. As noted in Sect.~\ref{sect:subcon}, this component shows
various signatures of ongoing dynamical activity, as the decentering of its brightest 
galaxy and the presence of interacting galaxies \citep{2011A&A...528A.115W, 2022hst..prop17212W}. 
This may result in excess star formation in this component. 
In the third component of A1185 the percentage of passive galaxies with old stellar 
populations is also very low.

The results on galaxy properties in clusters and their regions of influence for A1656 and A1367 
agree with the finding by \citet{2021A&A...649A..51E} in their study of the Corona Borealis (CB) supercluster.
They find a certain concordance between 
galaxy clusters and their close environment: clusters with a lower fraction of quiescent 
galaxies also have a lower fraction of such galaxies
among infalling groups and filaments. 
Clusters A1367 and Gr2064 in the CB also have the same total number of filaments connected to them, 
$N_{\mathrm{fil}}^{all} = 6$ (Table~\ref{tab:clresults}). However, A1656
also has six filaments and a larger number of substructures while having a lower percentage of
star-forming galaxies in its neighbourhood. The cluster A1185 has an even lower percentage of 
quiescent galaxies in the main component of the cluster than in its sphere of influence. 
This shows a large variety in galaxy populations of clusters and around them.

The difference between galaxy populations in high (excluding rich clusters) and 
low global density environments is also highly significant. 
It is interesting to note that galaxies with the oldest stellar populations
($D_{n}(4000) \geq 2.0$) reside also in the low-density environments populated mostly by poor groups
and single galaxies. 
We tested whether this result changes if we change  the threshold density
$D8$ used to separate high and low global density regions. We compared galaxy content
of high and low global density environments using global density range $4 \leq D8 \leq 6$.
This changes the number of galaxies in high- and low global density environments, and 
also changes the median values of $D_{n}(4000)$ index and $\log \mathrm{SFR}$, but this 
change is very small, less than $1$~\%, and statistical significance of the differences 
remains very high, with $p << 0.01$. This agrees well with the much more detailed
comparison of the galaxy content of various global environments in \citet{2022A&A...668A..69E}.

However, galaxy populations in 
filaments and  outside of filaments are 
statistically similar. 
This similarity does not change if we change the criteria of filament membership
(distance of a galaxy to the filament axis, to be a member of a filament).
Such similarity was noticed also in \citet{2024MNRAS.534.2228P} 
who analysed the effect of filaments of galaxies from IllustrisTNG simulations. 
One reason for this similarity may be that filaments may cross regions of
various global densities. Long filaments may extend from high-density cores of superclusters 
to low-density regions (voids) and may be populated differently along the filaments
\citep[see also corresponding discussion in ][]{2020A&A...641A.172E}.

We emphasise that in total, quiescent galaxies outside clusters
in the low global luminosity-density regions outnumber almost ten times the same population
in three richest clusters in our study (Fig.~\ref{fig:viogal} and Table~\ref{tab:clgal}). 
Even in the low-density region among grouped and single galaxies 
$25$~\% have very old stellar populations, an indicator of preprocessing
of these galaxies far from rich clusters.  The same was shown in \citet{2022A&A...668A..69E},
who discussed the possible mechanisms of galaxy quenching in poor groups and
in those galaxies which do not belong to any group in low-density environment.

In Fig.~\ref{fig:massfil}, we compare cluster masses, substructure, connectivity, and galaxy content,
and show the trend that clusters with a higher mass also have
a larger number of filaments connected to them, and a larger number of components.  
The cluster with the highest mass in our sample, A2065 in the Corona Borealis supercluster
($ID = 4$),  
has the largest total number of filaments 
connected to it. From Table~\ref{tab:clresults} we see that it also has the highest
mass embedded in it's sphere of influence.
Corresponding values for A1656 (cluster 1 in Fig.~\ref{fig:massfil})
are close to those for the cluster A2142 (cluster 8).
Our mass estimates approximately agree with those from the constrained simulations
of the local Universe \citep{2024A&A...687A.253H}. 
Figure~\ref{fig:massfil} also shows variety in cluster properties -- one of the clusters with a 
higher percentage of star-forming galaxies (A1367, cluster 2) 
has small values of substructure and connectivity, while A1185 (cluster 3) 
has a large number of components, but three long filaments connected to it.

\subsection{Scaling relations for masses and radii of clusters}
\label{sect:scaling}  

To find possible scaling relations between different masses of clusters and their environment,
and between various radii of clusters, we present in Fig.~\ref{fig:massmass}
the masses of clusters $M_{\mathrm{cl}}$ versus mass in the regions of influence ($M_{\mathrm{inf}}$,
left panel), and
in the turnaround region ($M_{ta}$, right panel). In Fig.~\ref{fig:massmass} symbols sizes 
are proportional to the total number of filaments connected to a cluster,
and colours indicate the percentage of quiescent galaxies in a cluster, as determined in
Sect.~\ref{sect:galcont}. 
In Fig.~\ref{fig:rgrrr} we show the relations between various radii of clusters and their
environment; the cluster gravitational radii $R_g$ and radii $R_\mathrm{cl}$, $R_\mathrm{inf}$, and 
$R_\mathrm{ta}$.
Figure~\ref{fig:massmass} shows tight correlation between cluster masses and 
the mass embedded in the region of influence and in the turnaround
region. 
For all clusters the mass in their spheres of 
influence is $M_{\mathrm{inf}} \approx 1.6 M_{\mathrm{cl}} + 0.2$, for all masses and connectivities.
The turnaround mass $M_{ta} \approx 2.2 M_{\mathrm{cl}} + 0.3$. 

The scaling relations between different mass estimates of 
clusters were presented by \citet{2025A&A...698A.272M}. In comparison with our work, there are important differences.
\citet{2025A&A...698A.272M} determined the scaling relations between cluster masses $M_{200}$
(the radius within region in which the mean density is 200 times the critical density of the universe),
and $M_{Hern}$, the mass calculated using Hernquist profile \citep[see][for details]{2025A&A...698A.272M},
in other words, within the same dark matter halo. They detected linear scaling relations in the case of clusters without significant substructure.
In our calculations regions of influence  and turnaround regions correspond to the environment
of clusters beyond the cluster dark matter haloes, and all clusters have substructure. 
The sizes of the regions
of influence and turnaround regions are larger for more massive clusters which represent 
deep potential wells, and embed
larger number of galaxies, groups, and filaments (Fig.~\ref{fig:massfil} and Fig.~\ref{fig:rgrrr}). 
Groups near more massive clusters  may also be of higher mass than groups near less massive 
clusters \citep[environmental enhancement of galaxy groups, see, for example,][]{2003A&A...401..851E}.
Therefore, positive correlations between
masses are expected. However, it is interesting that the scatter in the relations between masses is so small, as the structures
surrounding each cluster are different. The regions of influence around A1656 and A1185 are dominated
by filaments while A1367 is surrounded by a large number of galaxy groups. The scatter of the scaling relations between
radii in Fig.~\ref{fig:rgrrr} is larger. Thus, before making conclusions we need to study larger sample
of clusters. 
Our findings 
extend the scaling relations between different mass estimates of galaxy clusters to a larger radii
than in previous studies.

\section{Discussion}
\label{sect:discussion} 

We showed that all three clusters in our sample have substructure,  
with $2 - 5$ components and up to six filaments connected to them. Substructures found in our study
agree well with parent clusters from the DESI cluster catalogue. 
More massive clusters have a higher number of components and
higher connectivity, but the scatter is large. 
We also found the characteristic radii of clusters, 
$R_{\mathrm{cl}}$, $R_{\mathrm{inf}}$, $R_{\mathrm{ta}}$, and $R_{\mathrm{FC}}$,
and the distribution of mass around clusters, which gives us the density contrast at the borders
of the regions of influence of clusters, $\Delta\rho_\mathrm{inf} = 50 - 60$, as well as
the sizes and masses of turnaround and future collapse regions. 
According to the star formation properties, A1656 and its region
of influence contain the lowest
percentage of star-forming 
galaxies among clusters under study.
Especially high is the percentage of star-forming galaxies in cluster A1185,
and in the region of influence of the cluster A1367. This percentage is also higher
than in and around comparison clusters from the A2142 and the Corona Borealis superclusters.
The scaling relation between cluster masses and 
the mass embedded in the region of influence and in the turnaround
region has a very small scatter. 
Below, we discuss what our results tell us about the evolution of clusters.

\subsection{Evolution of clusters: From formation to future}
\label{sect:subdyn}  

{\emph{Cluster formation and virialisation}}. 
Simulations show that the formation of structures in the universe -- galaxies,
groups, and clusters is modulated by the combination of
density waves. Galaxies and their systems can form in the density field where 
positive phases combine so that the combination of small- and large-scale
density perturbations is sufficiently high
\citep{2011A&A...534A.128E, 2011A&A...531A.149S, 2021arXiv210602672P}.
The richest clusters form at the highest combined density peaks and grow by the infall
of galaxies and groups along filaments.
Based on zoom-in simulations of galaxy clusters (THE THREEHUNDRED project) 
\citet{2022MNRAS.510..581K} estimated that up to $45$~\% of galaxies fall into clusters
along filaments. 
\citet{2013ApJ...779..127C} estimate that protoclusters of Coma-like clusters with present-day mass
$M_{z0} \approx 1.0\times~10^{15}\mathrm{M}_\odot$
increased their masses ten times during evolution from $z = 2$ (approximately $10$~Gyr ago)
to the present day. 
An effective radius of such clusters was $R_\mathrm{eff} \approx 7 - 8$~\Mpc\ at redshift $z = 2$. 
The masses of protoclusters  with present-day mass
$M_{z0} \approx 0.5\times~10^{15}\mathrm{M}_\odot$ (as A1367 and A1185)
were approximately five times lower at  $z = 2$, and their effective radius at $z = 2$ was
$R_\mathrm{eff} \approx 5$~\Mpc. 
The actual growth of a particular protocluster depends on its surrounding structures. 

Based on the density contrast evolution, the turnaround of the virialised central parts
of clusters in our study with  $\Delta\rho = \Delta\rho_\mathrm{vir}$ at present
occured approximately at redshift $z \approx 0.7$ (Fig.~\ref{fig:vir-dens}) or
when the age of the Universe was  $7.5$~Gyr. 
At that time, in A1656 galaxies with $D_{n}(4000) \geq 2$ 
already stopped forming stars. 
The  percentage of such galaxies is the highest in the 
main component and in the third component of A1656.
Such galaxies formed their stellar populations approximately $10$~Gy ago, at redshift $z \approx 1.5 - 2$. 
Other components have lower $D_{n}(4000)$ and higher star formation rate values. 
The W direction from A1656 with its substructures and filaments pointing towards A1367 
is probably the main direction for cluster growth for A1656. 
 This interpretation agrees 
with the position  of components and filaments in  the 
PPS diagram (Fig.~\ref{fig:clskyppsdrho}), as well as with the predictions from simulations and observations 
of the evolution of protoclusters, and the brightest cluster galaxies 
\citep{2013ApJ...779..127C, 2025AJ....169..285L, 2025MNRAS.541..409H}.

In other clusters the percentage of star-forming
galaxies is  higher than in A1656 at present, 
and we may assume that it was higher also at redshifts $z \approx 0.7$. 
The PPS diagram shows that the second component of A1367 lies in the late infall region.
Figure~\ref{fig:skygr15wh24} shows that in the surroundings of A1367 there is more galaxy groups
than in the neighbourhood of A1656, with the higher percentage of star-forming galaxies in its region
of influence than in the cluster or in the region of influence of A1656 
\citep[groups and galaxy populations in A1367 were also described in][]{2020MNRAS.497..466S}. 
We may assume that with its lower mass, the infall of surrounding groups
has been slower in this cluster, which leads to longer timescales of cluster formation and group
merger times in comparison with
A1656, and
to ongoing star formation activity in the cluster and its region of influence.

In the cluster A1185 the main component has lower percentage of passive galaxies 
with old stellar populations than another, smaller component.
This may be related to the clumpy distribution
of passive galaxies in the main component, a signature of 
dynamical activity in this component, seen also in the X-ray maps
of the cluster. This activity 
enhances the star formation in galaxies \citep{1996AJ....111...64M, 
1985ApJ...293...94H, 2011A&A...528A.115W, 2012ApJS..199...22H}.
However, \citet{2025A&A...694A.216J} determined many small subclumps also in the
A1656, while the percentage of star-forming galaxies in Coma is lower than in A1185.
The third component in A1185 with the highest percentage of
star-forming galaxies and the youngest stellar populations 
may represent a small group falling into the cluster for the first time. 

Clusters, especially A1656 have galaxies with the oldest stellar populations in their
main component (early infall region). Recently, \citet{2025MNRAS.541..409H}
showed that in the GOGREEN and GCLASS surveys at redshifts $0.8 < z < 1.5$,
cores of clusters have higher fraction of quiescent galaxies than outskirts of clusters,
in agreement with findings in our study. \citet{2025MNRAS.541..409H} suggest that
in the early infall (core) region of clusters galaxy quenching is mostly related to accelerated 
mass dependent quenching of galaxies in protoclusters, while in late infall (non-core) regions 
quenching is better described by environmental, mass-independent processes of 
infalling galaxy populations.

{\emph{Regions of influence}}. 
Present-day regions of influence were next regions around clusters to reach turnaround 
and to start collapse. 
We found that the radii of the regions of influence of clusters are 
$R_\mathrm{inf} \approx 4$~\Mpc, and the density contrast at the borders 
of these regions is $\Delta\rho_\mathrm{inf} \approx 40 - 70$. 

The relation between the present (at $z = 0.02-0.03$) density contrast and the turnaround redshift 
$z_\mathrm{ta}$ at a specific redshift
is presented in Fig.~\ref{drhota}. We can see that if the 
present density contrast at the borders of regions of influence is
$\Delta\rho_\mathrm{inf} \approx 40 - 70$, then the turnaround  should have occurred at  
redshifts $z_\mathrm{ta} \approx 0.47 - 0.57$.
In the Corona Borealis and A2142 superclusters (at $z = 0.07 - 0.09$)  
$\Delta\rho_\mathrm{inf} \approx 30 - 40$, which means that 
regions of influence of
rich clusters in these superclusters 
were at the turnaround and started to collapse at redshift $z_\mathrm{ta} \approx 0.48-0.55$ 
\citep{2020A&A...641A.172E, 2021A&A...649A..51E}. 
Figure~\ref{fig:vir-ta} suggests that if the turnaround occurred at the redshifts
$z_\mathrm{ta} \approx 0.4-0.6$ then such regions will virialise in the future, 
approximately after $3.3$~Gyr from now (at the age of the Universe  $\approx 17.1$~Gyr, 
corresponding to $z \approx -0.2$ in Fig.~\ref{fig:vir-ta}).

{\emph{Turnaround and  future evolution}}.
The current radii of turnaround regions around A1656 and  A1367 are 
$R_{\mathrm{ta}} \approx 7 - 8$~\Mpc, and the radii of the future 
collapse regions are $R_{\mathrm{FC}} \approx 8 - 10$~\Mpc\ (Table~\ref{tab:clresults}).  
Mass, embedded in these regions are of order of
$M_{\mathrm{ta}} \approx (1.5 - 2.5) \times10^{15}\mathrm{M}_\odot$,
and $M_{\mathrm{FC}} \approx (2.0 - 2.5)\times10^{15}\mathrm{M}_\odot$.
Based on Fig.~\ref{fig:vir-ta}, we can predict that the regions at turnaround now (at $z = 0$) 
will be virialised after $10$~Gyrs from now, at the age of the Universe 
$\approx 22.9$~Gyr.

Figure~\ref{fig:skygr15wh24} shows that the distance between these clusters is 
approximately $24$~\Mpc, with the minimum in the distribution of groups
between clusters approximately at $9$~\Mpc\ from A1656. Although
at these distances there is a maximum of single galaxy distribution, these galaxies
do not contribute much to the total mass
{\footnote{For example, \citet{2021A&A...649A..51E} found that approximately $5$\%
of the mass of the Corona Borealis supercluster comes from single galaxies. }.
Therefore, even if we underestimate the mass and size of the turnaround region 
around A1656 and around A1367, it is still unlikely that A1656 and A1367 will merge in the future to form 
massive, collapsing supercluster core. 
This conclusion is somewhat different made by 
\citet{2024PASA...41...78Z}, who proposed that the clusters A1656 and A1367 perhaps may 
collapse in the future into one system.
To merge in the future, the size of the future collapse regions around clusters
A1656 and A1367 should be at least $12 - 13$~\Mpc, so that these regions overlap
and can collapse together. This means that the mass in these regions should be 
at least $M = 5 - 10\times10^{15}\mathrm{M}_\odot$ (see Fig.~\ref{fig:massradius}). 
This mass is in the same 
order as the mass of a collapsing core of the Shapley supercluster which embeds 11 rich
clusters or in the Corona Borealis supercluster 
\citep{2015A&A...575L..14C,  2021A&A...649A..51E,2024A&A...689A.332A}. Also, this leads
to 5--10 times higher mass-to-light ratio of these clusters, up to $M/L \approx 3000 - 4000$,
which is highly unlikely \citep[see also discussion of mass-to-light ratios
and the size of collapsing regions in][]{2015A&A...580A..69E, 2022A&A...666A..52E}.

The cluster A1185 is far away from other clusters in our sample, and it will
collapse separately. Its future collapse size and mass
are of order of  $R_{\mathrm{FC}}^{A1185} 8.1$~\Mpc\, and 
$M_{\mathrm{FC}}^{A1185} \approx 1.8\times10^{15}\mathrm{M}_\odot$ (Table~\ref{tab:clresults}).  
It is interesting to note that although the environments of both A1656 and A1185 are dominated
by filaments, their evolution and present dynamical state is different, as
suggests the galaxy content of these clusters and their regions of influence.

\citet{2014A&A...562A..87E} and \citet{2017ApJ...835...56C} showed
that in superclusters where galaxy clusters are 
connected by a large number of filaments (superclusters of (multi)spider type) the fraction of 
star-forming galaxies is higher, and clusters have more substructures than in superclusters
with simple structure (small number of filaments between clusters).
In agreement with this, \citet{2024ApJ...976..154K} found that clusters with larger number
of FoF-connected structures in their neighbourhood tend to have higher fraction of
star-forming galaxies at  redshifts approximately $0.4$ and higher, up to redshifts
$z \approx 0.9$. At still higher redshifts clusters have high fraction of star-forming
galaxies no matter how large is the number of connected structures around them - 
the properties of clusters in the IllustrisTNG simulation at redshift $z = 1$ 
are quite homogeneous.

Clusters grow by infall of matter from surroundings, 
and simulations enable their evolution to be followed  back 
up to very high redshifts \citep{2013ApJ...779..127C, 2017ApJ...844L..23C}.
Masses of massive clusters have grown more than twice
since redshift $z = 1$ \citep{2015JKAS...48..213K}.
The properties of clusters in the IllustrisTNG simulation at redshift $z = 1$ 
are quite homogeneous \citep{2024ApJ...976..154K}.
We found that the properties of clusters in our sample show large variety
in substructure, connectivity, and galaxy content. We may assume that the origin of this variety
is related to the evolution of clusters from $z = 1$ to the present.  
This shows the need for a further study of a large sample of clusters
both from observations and simulations.

\subsection{Cluster masses, connectivity, and substructure}
\label{sect:massconsub} 
In Fig.~\ref{fig:massfil} we showed large variety in the connectivity of clusters of the same mass, 
or, in other words, clusters of the same connectivity and/or the same number of components 
may have very different masses. We can compare this result with the masses, dynamical state, and connectivity of clusters in IllustrisTNG300 simulations by \citet{2021A&A...651A..56G}. 
\citet{2021A&A...651A..56G} 
found that at the same cluster mass, relaxed clusters from Illustris
TNG300 simulations have lower connectivity determined using T-ReX filament finder, than unrelaxed clusters. 
We get the same trend, but there are differences. 
In  our study, the scatter in connectivity values is larger than found in \citet{2021A&A...651A..56G},
and we apply Bisous filament finder \citep{2014MNRAS.438.3465T, 2016A&C....16...17T}. 
We found that the connectivity values, when using all filaments, are $C = 2 - 9$, and 
for long filaments connectivity $C = 1 - 5$, versus $C = 3 - 4$ in \citet{2021A&A...651A..56G}). 
Also, in \citet{2021A&A...651A..56G} connectivity is higher for clusters with substructure (nonregular clusters). 
In our study all clusters have substructure, and masses of clusters are higher than in IllustrisTNG300 simulations. 
Partly the differences between the studies  come from different definition of filaments and connectivity.

Theoretical predictions in \citet{2018MNRAS.479..973C} showed that the connectivity of massive haloes is approximately 5.
The number of long filaments in our study is lower than that, but the number of all
filaments, on average, agree with this prediction.  
These examples show the same trends between masses, connectivity, and substructure using different estimates of the substructure
and connectivity of clusters, but also emphasise that direct comparison of different
studies may not be straightforward, as the results depend on the methods applied to
determine connectivity and substructure. Also, if  filaments are connected 
to a cluster from both sides, then in our study these are considered as different filaments,
but in some other studies - as the same filament \citep[see also][]{2018MNRAS.479..973C}.
The comparison of cluster masses, substructure and connectivity from observations
and simulations is yet to be done in the future studies.

\subsection{Large-scale environment of galaxies}
\label{sect:large} 
In the supercluster outskirt regions (in superclusters, excluding rich clusters, thus
in poor groups and clusters and among single galaxies), there is approximately ten times 
more quiescent galaxies 
than  in clusters. 
In the whole superclusters ($D8 > 5$) the percentage of quiescent galaxies is higher 
 than in low-density regions between superclusters ($D8 < 5$), 
 with median values $D_{n}(4000)_{\rm med} \approx 1.52$ in superclusters 
 and $D_{n}(4000)_{\rm med} \approx 1.40$ in low-density region between superclusters 
 (see Table~\ref{tab:clgal}, for $D8 > 5$ and $D8< 5$, respectively).
\citet{2025A&A...700A..68S} found that quiescent galaxies at $z = 3$ are similar in various
environments (protoclusters, filaments). They concluded that 
quenching mechanisms are likely driven by similar physical processes independent of their environments. 
This is the same conclusion as in \citet{2022A&A...668A..69E, 2024A&A...681A..91E, 2025A&A...693L..16B}:
quenching occurs in all environments, and in low-density environments the number of 
quenched galaxies is even larger than
in rich clusters (Table~\ref{tab:clgal}, the last line).

\citet{2020A&A...635A.195G} analysed galaxy content in clusters and filaments from Magneticum hydrodynamical simulations
and found that passive galaxies trace filamentary pattern around clusters better than star-forming galaxies.
This suggests that quenching mechanisms are related to processes within galaxies and their 
dark matter haloes or in their immediate neighbourhood (infalling filaments), like AGN feedback, 
morphological quenching, detachment of primordial filaments and others are more important
in quenching galaxy formation in galaxies in low-density environments
than environmental processes related to
galaxy mergers and interactions in high-density environments.
Galaxy quenching via detachment of primordial filaments (Cosmic Web Detachment,
CWD) have been described in detail by \citet{2019OJAp....2E...7A}. 
They show that  
the accretion of gas into galaxies from primordial gaseous filaments around them
stops when these filaments are disrupted due to the influence of other galaxies, infall to groups
and so on.
This process leads to the end of
star formation in galaxies, combining several  mechanisms such as  starvation, 
harassment, and strangulation. 
\citet{2022A&A...668A..69E} suggested that the CWD may be
one of the mechanisms of galaxy quenching in global low-density environments.
\citet{2021MNRAS.501.4635S} apply DisPerSE 
filament finder to galaxies from HORIZON-AGN simulations
and find that the star formation in galaxies is suppressed at the edge of filaments.
They suggest that gas transfer to haloes becomes less efficient closer to filaments
which may explain the higher percentage of passive galaxies in filaments.

\section{Summary}
\label{sect:sum} 

We have studied the substructure, connectivity,  galaxy content,
characteristic radii, and mass distribution 
of the richest clusters in the Coma and Leo superclusters, A1656, A1367, and A1185
and their environments. We used these data to analyse  the timeline of the formation 
and evolution of these clusters and superclusters. We summarise our results as follows:

\begin{itemize}

\item[1)]
The clusters under study have
two to four components and a connectivity of $C = 5 - 6$. 
The highest percentage of quiescent galaxies is in A1656,
and the lowest is in A1185. In the region of influence of A1185, the
percentage of quiescent galaxies is higher than in the cluster itself, due to the
high percentage of quiescent galaxies in an infalling group.
\item[2)]
Outside of the clusters, in the superclusters and in the low-density region between superclusters
with no rich groups, 
approximately 25\% of the galaxies are quiescent. They are mostly located in small groups,
or they are single galaxies. As the richest clusters contain only a small fraction of all galaxies, 
quiescent galaxies outside of rich clusters
outnumber such galaxies in clusters by at least a factor of ten.
\item[3)]
The radii of the regions of influence of all clusters is $R_{\mathrm{inf}} \approx 4$~\Mpc, 
and the density contrast is $\Delta\rho_{inf} \approx 50 - 60$.
This suggests that the turnaround of
the regions of influence occurred, and the collapse started at redshifts of
$z \approx 0.4 - 0.5$. 
The turnaround regions around clusters will virialise in the future, approximately 
$9-10$~Gyr from now.
\item[4)]
In the future, these clusters will not merge; they will form separate 
collapsing structures with radii of $R_{\mathrm{FC}} \approx 8 - 10 $~\Mpc 
and masses of  $M_{\mathrm{FC}} \approx (1.8 - 2.5) \times10^{15}\mathrm{M}_\odot$.
\item[5)]
Our study extends the known scaling relations between cluster masses 
to larger distances from clusters.
The scaling relations between the cluster mass and the mass
embedded in the regions of influence and in the turnaround regions
have a very small scatter. 
\end{itemize}

The properties of clusters  (their substructure, connectivity, and
galaxy content) vary strongly, which suggests that they each experienced a different evolution.
Our findings indicate the self-similarity of mass distribution around clusters and 
extend the scaling relations between different mass estimates of galaxy clusters to a larger radii
than in previous studies.
\citet{2024ApJ...976..154K} showed that at redshift $z = 1$ and higher,
the properties of clusters in the IllustrisTNG300 simulation 
are quite homogeneous. 
The scatter of cluster connectivities in the IllustrisTNG simulation is lower than 
found in our study \citet{2021A&A...651A..56G}.
According to Horizon Run 4 simulations, the clusters with a present-day
mass of order of $M \approx \times~10^{15}M_\odot$ have gathered half of their 
masses between redshifts 1 and 0 \citep{2015JKAS...48..213K}.
This rises
the question as to whether the large variety of present-day cluster properties appeared during 
the evolution in last $7 - 8$~Gyrs or if there is a difference between observations and
simulations. 
These questions point to 
the significance of conducting future studies that analyse a large number of clusters, their environments,
connectivity, and galaxy properties simultaneously using both observations and large simulations.
Such studies will involve the use of data from present and forthcoming
surveys such as the J-PAS survey \citep{2014arXiv1403.5237B} and especially the forthcoming
4MOST survey \citep{2019Msngr.175....3D, 2019Msngr.175...46D, 2023Msngr.190...46T} 
and data from multiwavelength studies. 
In the future, the 4MOST cluster survey will increase the number 
of clusters and supercluster regions that can be studied in detail,
including the environment of clusters up to $5R_{200}$ \citep{2025A&A...697A..92S}.

\begin{acknowledgements}
We thank the referee for valuable comments and suggestions
which helped us to improve the paper.
We thank Hyeong Han Kim and Mirt Gramann for discussions.
This work was supported by the Estonian Ministry of Education and Research (grant TK202,
 “Foundations of the Universe"), 
Estonian Research Council grant PRG1006, by Estonian Research Council grants PRG2172 and PRG2159, and the 
European Union's Horizon Europe research and innovation programme 
(EXCOSM, grant No. 101159513).
This work has also been supported by
ICRAnet through a professorship for Jaan Einasto.
We applied in this study R statistical environment 
\citep{ig96}, and Julia language \citet{bezanson2017julia}.

\end{acknowledgements}

\bibliographystyle{aa}
\bibliography{comaleo.bib}

\begin{appendix}

\section{Spherical collapse and virialisation model} 
\label{sect:sph} 

Let us have a spherical overdensity $\Delta\rho \equiv \rho / \rho_m$ within a radius $R$,
where $\rho$ is the matter density in the volume, and $\rho_m$ is the mean matter density. 
As the universe expands,
the overdensity expands also. In principle, some matter falls to the overdensity, 
but it is not dominating, and usually, the mass of the overdensity is assumed to be constant 
(as we also do). 
The growth of masses are large at the formation stages of overdensities at very high redshifts 
($z > 20 - 30$) but such high redshifts are beyond our analysis.

In overdensity regions, due to internal gravitation, expansion is slower than in the 
surrounding background.
At a particular moment, the (slower) expansion may stop, and the region starts to contract. 
This is called a turnaround epoch. Depending on a value of the density contrast $\Delta\rho$, 
the turnaround may occur in the past, present, or future or may not occur at all. We can calculate 
these ''critical'' density contrasts for a given cosmological model. 

Let us now fix our cosmological model. We assume an expanding $\mathrm{\Lambda}$CDM universe. 
Cosmological parameters are taken from \citet{2011ApJS..192...18K}:
present matter density and dark energy density parameters $\Omega_{m0} = 0.27$ and $\Omega_{\Lambda 0}=0.73$, respectively. The Universe is assumed to be flat $\Omega=1$ with dark energy parameter $\omega = -1$. 
In our cosmological model, we do not look at a very early time. Thus, we need not to consider radiation; the matter is only baryonic and dark matter. The density of the dark energy is assumed to be constant in space and time. 

The time evolution of the matter density parameter is 
\begin{equation}
\Omega_m (z) = \frac{\Omega_{m0} (1+z)^3}{\Omega_{m0}(1+z)^3 + \Omega_{\Lambda 0}}
\label{eq:omega}
\end{equation}
and of the Hubble function is 
\begin{equation}
H(z) = H_0\sqrt{\Omega_{m0} (1+z)^3 + \Omega_{\Lambda 0}}.
\end{equation}

The scale factor of the universe is designated as $a$. In following we use standard designations: overdensity is $\Delta\rho \equiv \rho / \rho_m$,  turnaround redshift is $z_{ta}$ and overdensity at turnaround is $\Delta\rho_{ta}$. Density contrast at an arbitrary time is 
\begin{equation}
\Delta\rho = \Delta\rho_{ta} \, \left( x/y\right)^3, \label{eq:den}
\end{equation}
where $x=a/a_{ta}$ and $y=R/R_{ta}$ are the scale factor and density contrast radius in turnaround units. Time, in this case, is also in dimensionless units 
\begin{equation}
\tau = H (z_{ta})\sqrt{\Omega_m(z_{ta})}\, t
\label{eq:time}
\end{equation} 
\citep[see, for example,][]{wang:1998}. 

Differential equations describing the time evolution of $x$ and $y$ were derived by \citet[][Appendix A]{wang:1998} 
and \citet{2010JCAP...10..028L}. The analytical solution of the equation for $x$ in the form of a hypergeometric function is well-known
\begin{equation}
\frac{2}{3} x^{3/2} F\left[ \frac{1}{2}, \frac{1}{2}, \frac{3}{2}, -\frac{x^3\rho_{\Lambda}}{\rho_{ta}} \right] = \tau .
    \label{eq:x}
\end{equation}
Here $F$ is the hypergeometric function. The solution of the equation for $y$ can be written as an integral, with integration constants derived from the boundary conditions at the turnaround (Eqs. 2.13 and 2.20 
in \citet{2010JCAP...10..028L}). Parameter $\Delta\rho_{ta}$ in these equations can be calculated then from the condition that $y(0)=0$ for different $z_{ta}$ values. We calculated corresponding integrals numerically and derived $\Delta\rho_{ta}$ values. Although an analytical approximation formula was proposed by 
\citet{2010JCAP...10..028L} we calculated more precise $\Delta\rho_{ta}$ values ourselves. 

In this way, for a fixed turnaround redshift $z_{ta}$ we have corresponding density contrast $\Delta\rho_{ta}$ and functions $x (\tau )$ and $y(\tau ).$ Corresponding density contrasts as a function of $\tau$ (and thereafter of $z$) result directly from Eq.~\ref{eq:den}. If density contrast is expressed relative to local matter density, then
\begin{equation}
\Delta\rho = \frac{\rho}{\rho_m} = \frac{\rho}{\rho_{m0}} (1+z)^{-3} = 0.860\cdot 10^{-12} \Omega_{m0}^{-1} (1+z)^{-3} M/R^3,
\label{eq:den_red}
\end{equation}
where $M$ is in units of $(h^{-1}\mathrm{M_\odot})$ and $R$ in $(h^{-1}\mathrm{kpc})$. 

Let us turn now to virialisation. 
In the spherical collapse model, after the overdensity starts to contract, the final stage is the collapsed structure with a theoretical 
radius of zero---a singularity. In real, 
there exist several relaxation processes such as mixing of orbits, violent relaxation, 
and Landau damping,\footnote{Mixing: Fundamental frequencies of nearby orbits differ a little, 
and thus their positions in phase space start to differ more and more, ultimately reaching 
a quasi-stationary state. In the case of chaotic orbits, these orbits move away exponentially 
from each other and end up filling all the available space. Violent relaxation: 
A nonstationary potential changes particle energies  so that their energy distribution expands. 
Landau damping: An exchange of energies between the perturbation waves and particles causes an 
increase of kinetic energies of particles due to the damping of waves.} allowing us to avoid the singularity. 
Thus, the final stage should be a virialised state. 
It is usually postulated that in the contracting stage, the virialised state (equilibrium) 
is reached at the radius from which the virial theorem holds. 
We also use this to define the virial radius and designate it as $R_{vir}$. 
Using the virial theorem, it can be shown that in the case of the EdS universe, 
the virial radius is $ R_{vir} = 0.5\,R_{ta}$ \citep[see, for example,][]{1991MNRAS.251..128L}, 
giving the corresponding density contrast value 147 
(instead of the usual value 178, derived by allowing the system to contract to singularity). 

In the case of a nonzero cosmological constant ($\rho_{\Lambda} = {\rm const}$), 
it is necessary to include also the dark energy terms to the virial theorem 
equation \citep{1991MNRAS.251..128L, wang:1998}. 
In this case ($\rho_{\Lambda} = {\rm const}$), 
the dark energy does not virialise (only the matter does) but still acts as a background and 
contributes to gravitational potential. 
Potential energy is not a function of time and energy is conserved. 
Using again the variable $y$ as $R_{vir} = y_{vir}\, R_{ta}$, it can be shown 
\citep[see][]{maor-lahav:2005, wang:2006, 2010JCAP...10..028L} that the 
virial theorem gives us for $y$ (or more precisely for $y_{vir}$) the relation
\begin{equation}
y^3 + \frac{1}{4} \left( \frac{\rho_{ta}}{\rho_{\Lambda}} -2 \right) y - \frac{\rho_{ta}}{8\rho_{\Lambda}} = 0.
\label{eq:y-vir}
\end{equation}
A similar equation was derived earlier by \citet{1991MNRAS.251..128L, wang:1998} but for pressure-less dark energy ($\omega = 0$).

In the case of variable dark energy density ($\rho_{\Lambda} \ne {\rm const}$), virialisation is much more complicated. In this case, one should consider that the dark energy is in direct gravitational interaction with the matter (it is not a constant background any more), and its contribution to gravitational potential is more complicated. Energy is not conserved also. In most general form, this was studied by \citet{maor-lahav:2005} and \citet{chang:2018}. 

In the present paper, we assume a constant dark energy density as supported by current CMB data ($\omega = -1$ within the errors). In calculations, we fixed first a turnaround redshift $z_{ta}$. This gives us $\rho_{ta} \equiv \rho_m (z_{ta}) = \Omega_m(z_{ta}) \rho_{cr}(z_{ta})$ and we solved Eq.~\ref{eq:y-vir} to find $y_{vir}$ and corresponding dimensionless time $\tau_{vir}$ from Eq.~\ref{eq:time}. After that, we calculated $x_{vir}$ from Eq.~\ref{eq:x}. 
The density contrast corresponding to virialisation 
\begin{equation}
\Delta\rho_{vir} = \Delta\rho_{ta} \left( x_{vir} / y_{vir} \right)^3
    \label{eq:vir-dens}
\end{equation}
Virialisation time (and thereafter redshift) can be calculated again from Eq.~\ref{eq:time}. This is the density contrast at which the collapse stops (does not go to zero), 
and the overdensity is virialised due to the processes referred to above.

\begin{figure}
\includegraphics[width=80mm]{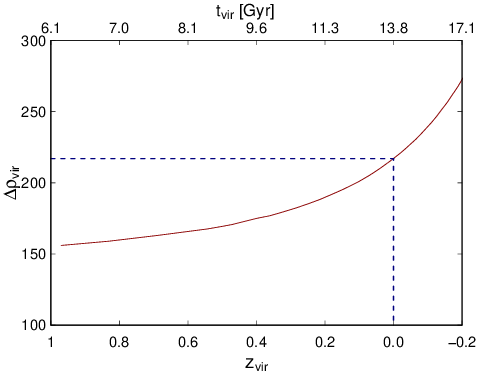}
\caption{Density contrast at the virialised epoch $\Delta\rho_{vir}$ as a function of corresponding redshift $z_{vir}$ (time in upper axis). Negative redshift values must be interpreted as future cosmological times, shown in upper axis.}
\label{fig:vir-dens}
\end{figure}

\begin{figure}
\includegraphics[width=82mm]{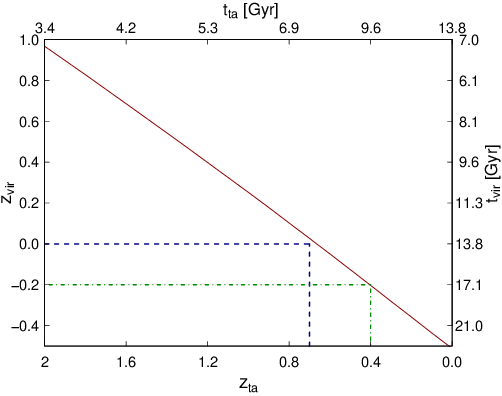}
\caption{Relation between the turnaround and virial redshifts,
$z_{ta}$ and $z_{vir}$. Negative redshift values must be interpreted as future cosmological times; 
timelines (corresponding age of the Universe) are shown in the right and upper axes. Systems 
which expansion have turned around at $z= 0.4$ (green dash-dotted line) or at $z = 0.7$ 
(blue dashed line) will be virialised in future (at 17.1 Gyr) or now
(at 13.8 Gyr).}
\label{fig:vir-ta}
\end{figure}

\begin{figure}
\includegraphics[width=80mm]{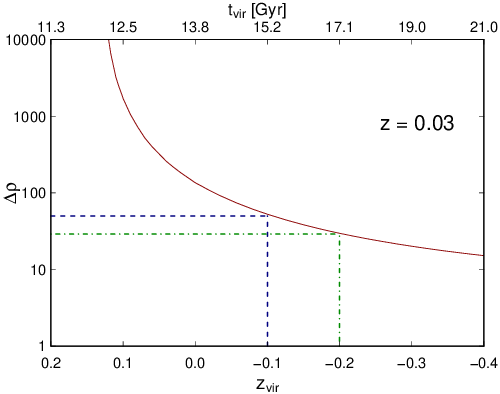}
\caption{Density contrasts $\Delta\rho = \rho /\rho_m$ at $z=0.03$ as a function of virialisation redshifts $z_{vir}$. Negative redshift values must be interpreted as future cosmological times 
(timeline is shown in the upper axis). Green dot-dashed line corresponds to the density contrast  $\Delta\rho = 30$,
and blue dashed line to the density contrast $\Delta\rho = 60$,
the range $\Delta\rho_{inf}$ found in this study.}
\label{f8}
\end{figure}
 
\begin{figure}
\includegraphics[width=80mm]{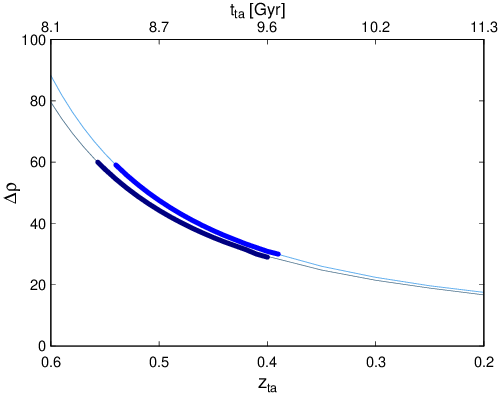}
\caption{Density contrast $\Delta\rho$ at redshift $z = 0.02-0.03$ (upper and lower curves, respectively) for $\Omega_\mathrm{m} = 0.27$,
if the turnaround occurred at redshifts in the interval $z = 0.6 - 0.2$. 
Thicker parts of the curves correspond to the density contrasts $\Delta\rho = 30-60$ (see Table 2 and Fig.~\ref{f8}).}
\label{drhota}
\end{figure}

Calculated density contrasts as a function of virialisation redshifts are shown in Fig.~\ref{fig:vir-dens}. We see that the density contrast corresponding to the 
virialisation in the local Universe ($z=0$) in the case of our used cosmological parameters is $\Delta\rho_{vir} = 217$. 
This is similar to the density contrast used in literature and corresponds to the radius $R_{200}$. However, we point out that the value is a function of redshift. For example, for systems at redshift $z=0.5$, all regions with $\Delta\rho > 169$ are 
virialised.

Virialisation redshift is related to turnaround redshift
\begin{equation}
	x_{vir}(z_{vir} + 1) = (z_{ta} + 1).\label{eq:vir-ta}
\end{equation}
Although, in principle, $x_{vir}$ is also a function of $z_{ta}$ this dependence is quite weak and from Fig.~\ref{fig:vir-ta} is is seen that Eq.~\ref{eq:vir-ta} is nearly a linear relation. Turnaround redshift corresponding to $z_{vir}=0$ is $z_{ta}=0.665$.

Now we can calculate density contrasts $\rho /\rho_m$ and densities $\rho /\rho_{m0}$ at any redshift, 
when the corresponding region is virialised at a certain redshift. 
In Fig.~\ref{f8} 
these values are given for a redshift $z=0.03$, 
corresponding approximately to the clusters from the present paper,
as a function of virialisation redshift. 
One can see that in the case of the present clusters at redshifts $z\sim 0.03$, all regions with density contrasts 
$\rho /\rho_m \ge 210$ have been virialised by the time we observe them. 
Corresponding value for the BOSS Great Wall is $\rho /\rho_m \ge 169$. 
We also calculated the density contrasts at our clusters redshifts
for cases in which the
turnaround occurs at redshifts in the range of $z = 0.6- 0.2$. 
In our calculations, we used formulae derived by \citet{2010JCAP...10..028L}.
We show the results for $\Omega_\mathrm{m} = 0.27$
in Fig.~\ref{drhota}.

\onecolumn

\section{Data on clusters: DESI clusters, and substructure}
\label{sect:apptabs} 

In this section we present data on clusters from WH24 catalogue (DESI data, Table~\ref{tab:wh24cl}),
and data on substructure of clusters, with identifications based on WH24m catalogue, and on
\citet{2024NatAs...8..377H} (Table~\ref{tab:sub}).

\begin{table*}[ht]
\caption{Data of rich galaxy clusters in the Coma and Leo superclusters identified in the WH24 cluster catalogue.}
\begin{tabular}{rrrrrrrr} 
\hline\hline  
(1) & (2) & (3) & (4) & (5) & (6) & (7) & (8) \\
   ID &  Name &  R.A. &  Dec. &  $z_{\rm cl}$ &  $r_{500}$ &  $M_{500}$ &  $N_{\rm gal}$  \\
\hline\hline  
\hline       
A1656 &&&&&&&\\
\hline       
1 & J130008.1+275837 & 195.0339 & 27.97697 & 0.0266 & 1.262 & 4.94 & 95 \\
2 & J125612.2+274444 & 194.0506 & 27.74549 & 0.0231 & 0.792 & 1.21 & 30 \\
3 & J125732.0+282837 & 194.3832 & 28.47696 & 0.0226 & 0.671 & 0.51 & 14 \\
4 & J130133.6+290750 & 195.3900 & 29.13058 & 0.0245 & 0.596 & 0.52 &  7 \\
\hline       
A1367&&&&&&&\\
\hline       
1 & J114402.2+195659 &176.0090 & 19.94982 & 0.0210 & 0.856  & 1.73 &28 \\
2 & J114612.2+202330 &176.5508 & 20.39164 & 0.0231 & 0.709  & 0.73 & 8 \\
3 & J114041.7+202035 &175.1736 & 20.34297 & 0.0218 & 0.628  & 0.71 & 6 \\
4 & J115242.6+203753 &178.1775 & 20.63131 & 0.0222 & 0.694  & 0.85 &11 \\
\hline       
A1185&&&&&&&\\
\hline       
1& J111038.4+284603 &167.6601 & 28.76758 & 0.0348 & 0.910 & 2.06 & 35 \\
2& J110718.2+283140 &166.8258 & 28.52776 & 0.0332 & 0.636 & 0.59 & 11 \\
3& J111203.3+273523 &168.0139 & 27.58977 & 0.0350 & 0.581 & 0.58 & 10 \\
\hline                                        
\label{tab:wh24cl}  
\end{tabular}\\
\tablefoot{                                                                                 
Columns are as follows:
(1): Cluster nr;
(2): Cluster name with J2000 coordinates;    
(3-4): Right Ascension (R.A. J2000) and Declination (Dec. J2000) of the cluster (in degrees);
(5): cluster redshift $z_{\rm cl}$;
(6): cluster radius, $r_{500}$, in Mpc; 
(7): cluster mass, $M_{500}$, in units of $10^{14}~M_{\odot}$;
(8): number of member galaxy candidates within $r_{500}$.
}
\end{table*}

\begin{table*}[ht]
\caption{Substructure of clusters.}
\begin{tabular}{rrrrrrrrr} 
\hline\hline  
(1) & (2) & (3) & (4) & (5) & (6) & (7) & (8) & (9) \\
   No. & $N_{\rm gal}$ &  R.A. &  Dec. &  $Dist_{\rm med}$ & $z_{\rm med}$ &  $log SFR_{med}$ & $D_{n}(4000)_{med}$ & ID \\
\hline\hline  
\hline       
A1656 &&&&&&&\\
\hline       
1 & 295 & 194.9 & 27.75 & 67.02 & 0.022 & -1.75 & 1.72 & 1;SE \\
2 &  60 & 194.8 & 28.89 & 73.24 & 0.024 & -1.52 & 1.67 & 3,4;N \\
3 & 119 & 195.5 & 28.10 & 63.75 & 0.020 & -1.75 & 1.72 & SE \\
4 &  48 & 193.2 & 27.37 & 70.84 & 0.023 & -1.62 & 1.64 & W \\
5 & 158 & 194.2 & 27.48 & 74.49 & 0.024 & -1.53 & 1.65 & 2,3;W \\
\hline       
A1367&&&&&&&\\
\hline       
1 & 178 & 176.1 & 19.84 & 65.34 & 0.021  & -1.54 & 1.60 & 1 \\
2 &  67 & 176.3 & 20.32 & 74.84 & 0.024  & -1.18 & 1.46 & 2 \\
\hline       
A1185&&&&&&&\\
\hline       
1 & 91 & 167.7 & 28.69 &  99.0 & 0.032 & -1.26 & 1.62 & 1  \\
2 & 23 & 168.0 & 27.57 & 105.1 & 0.034 & -1.53 & 1.76 & 3 \\
3 & 27 & 166.8 & 28.66 & 100.0 & 0.032 & -0.56 & 1.25 & 2 \\
4 & 60 & 167.7 & 28.26 & 108.5 & 0.035 & -1.01 & 1.57 &   \\
\hline                                        
\label{tab:sub}  
\end{tabular}\\
\tablefoot{                                                                                 
Columns are as follows:
(1): Component number;
(2): Number of galaxies in a component;    
(3-4): Median Right Ascension (R.A.) and Declination (Dec.) of galaxies in a component (in degrees);
(5): Median uncorrected distance of galaxies in a component, $Dist_{\rm med}$, in \Mpc; 
(6): Median observed redshift of galaxies in a component, $z_{\rm med}$;
(7): Median value of star formation rate of galaxies in a component, $log SFR_{med}$;
(8): Median value of $D_{n}(4000)$ index of galaxies in a component, $D_{n}(4000)_{med}$;
(9): identification of a component with structures from other studies. Numbers are numbers
of partner clusters from WH24m catalogue (Table~\ref{tab:wh24cl}). W, N, and SE denotes
endpoints of filaments as identified in \citet{2024NatAs...8..377H}.
}
\end{table*}

\twocolumn

\section{SDSS and DESI surveys}
\label{sect:surveys}

We are pleased to thank the SDSS survey team and DESI survey team for the publicly available data
releases.  Funding for the Sloan Digital Sky Survey (SDSS) and SDSS-II has been
provided by the Alfred P. Sloan Foundation, the Participating Institutions,
the National Science Foundation, the U.S.  Department of Energy, the
National Aeronautics and Space Administration, the Japanese Monbukagakusho,
and the Max Planck Society, and the Higher Education Funding Council for
England.
The SDSS\footnote{http://www.sdss.org/} is managed by the Astrophysical Research Consortium (ARC) for the
Participating Institutions.  The Participating Institutions are the American
Museum of Natural History, Astrophysical Institute Potsdam, University of
Basel, University of Cambridge, Case Western Reserve University, The
University of Chicago, Drexel University, Fermilab, the Institute for
Advanced Study, the Japan Participation Group, The Johns Hopkins University,
the Joint Institute for Nuclear Astrophysics, the Kavli Institute for
Particle Astrophysics and Cosmology, the Korean Scientist Group, the Chinese
Academy of Sciences (LAMOST), Los Alamos National Laboratory, the
Max-Planck-Institute for Astronomy (MPIA), the Max-Planck-Institute for
Astrophysics (MPA), New Mexico State University, Ohio State University,
University of Pittsburgh, University of Portsmouth, Princeton University,
the United States Naval Observatory, and the University of Washington.

The DESI Legacy Surveys consist of three individual and complementary projects: 
the Dark Energy Camera Legacy Survey (DECaLS; Proposal ID \#2014B-0404; PIs: David Schlegel and Arjun Dey), 
the Beijing-Arizona Sky Survey (BASS; NOAO Prop. ID \#2015A-0801; PIs: Zhou Xu and Xiaohui Fan), 
and the Mayall z-band Legacy Survey (MzLS; Prop. ID \#2016A-0453; PI: Arjun Dey). 
DECaLS, BASS and MzLS together include data obtained, respectively, at the Blanco 
telescope, Cerro Tololo Inter-American Observatory, NSF’s NOIRLab; the Bok telescope, 
Steward Observatory, University of Arizona; and the Mayall telescope, Kitt Peak National Observatory, 
NOIRLab. Pipeline processing and analyses of the data were supported by NOIRLab and the 
Lawrence Berkeley National Laboratory (LBNL). The Legacy Surveys project is honored to be 
permitted to conduct astronomical research on Iolkam Du’ag (Kitt Peak), a mountain with 
particular significance to the Tohono O’odham Nation.

NOIRLab is operated by the Association of Universities for Research in Astronomy (AURA) 
under a cooperative agreement with the National Science Foundation. LBNL is managed by the 
Regents of the University of California under contract to the U.S. Department of Energy.

This project used data obtained with the Dark Energy Camera (DECam), which was constructed 
by the Dark Energy Survey (DES) collaboration. Funding for the DES Projects has been provided 
by the U.S. Department of Energy, the U.S. National Science Foundation, the Ministry of 
Science and Education of Spain, the Science and Technology Facilities Council of the 
United Kingdom, the Higher Education Funding Council for England, the National Center 
for Supercomputing Applications at the University of Illinois at Urbana-Champaign, the 
Kavli Institute of Cosmological Physics at the University of Chicago, Center for Cosmology 
and Astro-Particle Physics at the Ohio State University, the Mitchell Institute for Fundamental 
Physics and Astronomy at Texas A\&M University, Financiadora de Estudos e Projetos, 
Fundacao Carlos Chagas Filho de Amparo, Financiadora de Estudos e Projetos, Fundacao 
Carlos Chagas Filho de Amparo a Pesquisa do Estado do Rio de Janeiro, Conselho Nacional 
de Desenvolvimento Cientifico e Tecnologico and the Ministerio da Ciencia, Tecnologia e 
Inovacao, the Deutsche Forschungsgemeinschaft and the Collaborating Institutions in the 
Dark Energy Survey. The Collaborating Institutions are Argonne National Laboratory, 
the University of California at Santa Cruz, the University of Cambridge, Centro de 
Investigaciones Energeticas, Medioambientales y Tecnologicas-Madrid, the University 
of Chicago, University College London, the DES-Brazil Consortium, the University of 
Edinburgh, the Eidgenossische Technische Hochschule (ETH) Zurich, Fermi National 
Accelerator Laboratory, the University of Illinois at Urbana-Champaign, the 
Institut de Ciencies de l’Espai (IEEC/CSIC), the Institut de Fisica d’Altes Energies, 
Lawrence Berkeley National Laboratory, the Ludwig Maximilians Universitat Munchen 
and the associated Excellence Cluster Universe, the University of Michigan, NSF’s NOIRLab, 
the University of Nottingham, the Ohio State University, the University of Pennsylvania, 
the University of Portsmouth, SLAC National Accelerator Laboratory, Stanford University, the 
University of Sussex, and Texas A\&M University.

BASS is a key project of the Telescope Access Program (TAP), which has been funded by 
the National Astronomical Observatories of China, the Chinese Academy of Sciences (the 
Strategic Priority Research Program “The Emergence of Cosmological Structures” Grant \# XDB09000000), 
and the Special Fund for Astronomy from the Ministry of Finance. The BASS is also supported 
by the External Cooperation Program of Chinese Academy of Sciences (Grant \# 114A11KYSB20160057), 
and Chinese National Natural Science Foundation (Grant \# 12120101003, \# 11433005).

The Legacy Survey team makes use of data products from the Near-Earth Object Wide-field 
Infrared Survey Explorer (NEOWISE), which is a project of the Jet Propulsion Laboratory/California 
Institute of Technology. NEOWISE is funded by the National Aeronautics and Space Administration.

The Legacy Surveys imaging of the DESI footprint is supported by the Director, 
Office of Science, Office of High Energy Physics of the U.S. Department of Energy 
under Contract No. DE-AC02-05CH1123, by the National Energy Research Scientific 
Computing Center, a DOE Office of Science User Facility under the same contract; 
and by the U.S. National Science Foundation, Division of Astronomical Sciences 
under Contract No. AST-0950945 to NOAO.

\end{appendix}

\end{document}